\documentclass{aa}  

\usepackage{graphicx}
\usepackage{txfonts}
\usepackage{lipsum}
\usepackage{subcaption}         
                                
\usepackage{lscape}            
                                
\usepackage{placeins}           
                                
\usepackage{orcidlink}

\usepackage{hyperref}
\hypersetup{
    colorlinks=true,
    citecolor=blue,
    linkcolor=blue,
    urlcolor=blue
    }

\usepackage{natbib}
\usepackage{tikz}
\bibpunct{(}{)}{;}{a}{}{,}

\begin{document}

   \title{X-ray stellar feedback in low-metallicity starbursts:}

   \subtitle{ Insights from the template starburst galaxy ESO 338-IG04 and its halo\thanks{Based on observations obtained with the science missions Chandra and XMM-Newton.}}

   \author{M. Chatzis\inst{1}\fnmsep\thanks{chatzis@astro.physik.uni-potsdam.de}\orcidlink{0009-0001-4052-7945} , L. M. Oskinova\inst{1}, S. Reyero Serantes\inst{1}\orcidlink{0000-0002-3155-7237}, 
   B.D. Lehmer\inst{2,3}, G. Östlin\inst{4}, A. Bik\inst{4}, M. Hayes\inst{4}\orcidlink{0000-0001-8587-218X}, J. M. Mas-Hesse\inst{5}\orcidlink{0000-0002-8823-9723}, J. S. Gallagher\inst{6}\orcidlink{0000-0001-8608-0408},
   F. Fürst\inst{7}
        }
\authorrunning{Chatzis et al.}
   \institute{Institute for Physics and Astronomy, University of Potsdam, Karl-Liebknecht-Str. 24/25, 14476 Potsdam, Germany
   \and 
   Department of Physics, University of Arkansas, 226 Physics Building, 825 West Dickson Street, Fayetteville, AR 72701, USA
   \and
   Arkansas Center for Space and Planetary Sciences, University of Arkansas, 332 N. Arkansas Avenue, Fayetteville, AR 72701, USA
   \and 
   Department of Astronomy, The Oskar Klein Centre, Stockholm University, AlbaNova, SE-10691 Stockholm, Sweden
   \and
   Centro de Astrobiolog\'{\i}a (CAB), INTA--CSIC, Madrid, Spain 
   \and
   Department of Astronomy, U. Wisconsin-Madison, 475 N. Charter St., Madison, Wisconsin 53706 USA
   \and
   ESA, ESAC, Apartado 78, 28691 Villanueva de la Cañada, Madrid, Spain
   }

   \date{Received September 30, 20XX}

  \abstract
   {The X-ray output of low-metallicity starburst galaxies is a key component of stellar feedback, tracing the processes responsible for gas ionization and chemical enrichment.
   The integrated X-ray luminosity ($L_X$) from high-mass X-ray binaries in star-forming galaxies scales with both the star formation rate (SFR) and the host galaxy's metallicity $Z$. 
   Due to the inverse correlation between $L_X$/SFR and $Z$, the contribution of X-ray binaries to the ionizing photon budget is expected to be enhanced in metal-poor systems.
   Their radiation can potentially ionize \ion{He}{ii} in the surrounding interstellar medium, powering nebular \ion{He}{ii}\,$\lambda 4686$\,\AA~emission.
   However, detailed studies of the X-ray emission in individual low-$Z$ starburst galaxies are rare, and their X-ray properties are not well explored.
   }
   {The blue compact dwarf galaxy ESO 338-IG04 (ESO 338-4 hereafter) serves as a nearby template for studying stellar feedback and X-ray emission in low-metallicity starbursts. 
   It combines vigorous recent star formation, low metallicity ($12+\log(\element{O}/\element{H})\approx7.9$, or $12\%$ solar), and a rich population of massive stellar clusters. 
   Extensively observed in optical and UV wavelengths with HST and VLT MUSE, ESO 338-4 is ideally suited for multi-wavelength feedback studies.
   In this work, we aim to characterize the X-ray emission of ESO 338-4 and its galactic halo using new deep \textit{Chandra X-ray Observatory (Chandra)} and \textit{XMM-Newton} observations. } 
   {We analyze X-ray spectra, lightcurves, and images of ESO 338-4 to constrain the nature of its X-ray sources.
   Additionally, we employ photoionization modeling to assess the significance of X-ray sources to the observed nebular \ion{He}{ii}\,$\lambda 4686$\,\AA~emission. }
   {We identify five ultra-luminous X-ray sources (ULXs) and diffuse hot gas surrounding ESO 338-4.
   Two of the ULXs are spatially associated with stellar clusters.
   The galaxy’s total X-ray luminosity exceeds $10^{41}$\,erg\,s$^{-1}$. 
   The brightest point source, ULX1, shows variability on timescales of days and is not associated with a stellar cluster.
   Lastly, our modeling demonstrates that X-ray sources significantly impact the galaxy’s ionizing photon budget.
   Photoionization modeling with ULX1 as the ionizing source predicts a high nebular \ion{He}{ii}\,$\lambda 4686$\,\AA~line luminosity of approximately $10^{39}\,\text{erg}\,\text{s}^{-1}$.
   
   }
    {}

   \maketitle
\nolinenumbers
\section{Introduction}

X-ray observations of metal-poor (low-$Z$) dwarf starburst galaxies are essential for understanding stellar feedback.
Their X-ray emission traces the energetic feedback from massive stars and supernovae that drives metal-enriched outflows into the circumgalactic medium. 
Yet, in contrast to the extensively studied hot halos of massive galaxies \citep{HalosMassive2024}, the X-ray properties of dwarf starburst halos remain poorly understood. 
Only a few detections have been reported to date \citep{Otherdwarfhalos2005,Matt_Haro11_2007}, and many low-mass galaxies do not have any detectable X-ray emission from their gaseous halos \citep{dwarfnotwellunderstood2025}.

The brightest X-ray point sources in young starbursts are high-mass X-ray binaries (HMXBs), which host a neutron star or black hole and accrete matter from massive donor stars through either stellar winds or Roche lobe overflow. 
The integrated X-ray luminosity from HMXBs scales with a galaxy's recent star formation rate (SFR) \citep{Grimm2003,Mineo2012}. The SFR-scaled X-ray luminosity function (XLF) also exhibits a dependence on gas-phase metallicity, producing a scaling relation between $L_X$/SFR and $12+\log(\element{O}/\element{H})$, with an observed inverse correlation \citep{OldXFL_bret2022,Lehmer2024}.
However, most observed galaxies have close to solar metallicities, highlighting the importance of careful studies of starbursts at the lowest metallicities.

An unresolved problem in studies of low-$Z$ galaxies is accounting for the total nebular \ion{He}{ii}\,$\lambda 4686$\,\AA~emission -- a central aspect of the so-called \ion{He}{ii} problem.
Models of the galactic ionizing photon budget that only include stellar populations often underpredict the observed \ion{He}{ii} luminosities \citep[e.g.,][]{Stasi2015,Stanway2019,Saxena2020a,Berg2021}.
The role of X-ray binaries in the ionizing photon budget remains debated: while several studies find their contribution insufficient to explain the observed \ion{He}{ii} luminosities \citep[e.g.,][]{Saxena2020b,Senchyna2020,Katz2023,Lecroq2024}, others demonstrate that they can supply a substantial fraction of the required ionizing photons \citep[e.g.,][] {Simmonds2021,OldXFL_bret2022,OskinovaScharer2022,Umeda2022,Garofali2024,KourSv2025}.

One of the best systems for studying the properties of low-metallicity starbursts is the blue compact dwarf galaxy (BCG) ESO 338-4.
Located at the distance of $40$\,Mpc \citep[HyperLEDA database,][]{HyperLedaDistance2014}, the galaxy has been extensively studied in the optical and UV with the Hubble Space Telescope (HST) \citep{Ostlin98,SSC_2003} and with the Multi Unit Spectroscopic Explorer (MUSE) instrument at the Very Large Telescope (VLT), as presented by the comprehensive review of \cite{Bik2018}.

\cite{Ostlin98,Ostlin2001,SSC_2003} investigated the age and masses of $124$ star clusters identified in HST UV and optical images, deriving that the present starburst has been active during the last $40$\,Myr, and showing evidence for propagating star formation and structures triggered by galactic winds. 
The SFR derived from their HST H$\alpha$  integrated luminosity was $3.2$\,M$_\odot$\,yr$^{-1}$ (corresponding to $3.6$\,M$_\odot$\,yr$^{-1}$ at the distance of $40$\,Mpc).
And later updated to $3.9$\,M$_\odot$\,yr$^{-1}$ 
($4.4$\,M$_\odot$\,yr$^{-1}$ at $40$\,Mpc) in \cite{Ostlin2009}.
This value is higher than the SFR of $1.9$\,M$_\odot$\,yr$^{-1}$ ($2.2$\,M$_\odot$\,yr$^{-1}$ at $40$\,Mpc) reported by \cite{Bik2018}, from their H$\alpha$ observations with MUSE.
This discrepancy arises from both the lower H$\alpha$ luminosity measured with MUSE and the different electronic densities assumed in the conversion to ionizing photons flux, as well as from differences in the adopted conversion factors: \cite{Ostlin2001} used their own calibration with a \cite{Salpeter1955} initial mass function (IMF), whereas \cite{Bik2018} follow the \cite{KennicuttEvans2012} prescription assuming a \cite{Kroupa2003} IMF. 
\cite{Lehmer2024}, in contrast, model the galaxy’s integrated spectral energy distribution from the far-UV to the far-IR and infer an SFR of $0.6$\,M$_\odot$\,yr$^{-1}$ averaged over the last $125$\,Myr. 

The spread in reported SFR values reflects both methodological differences and the distinct timescales probed by the various indicators, and thus represents the inherent method-dependent variance in SFR estimates for ESO 338-4.
In this paper we will assume the mean value of the SFRs derived from H$\alpha$ luminosities, $\text{SFR}=3.3$\,M$_\odot$\,yr$^{-1}$, when referring to short term phenomena as the diffuse soft X-ray emission, and the 125 Myr averaged value, $\text{SFR}_{125\,\text{Myr}}=0.6$\,M$_\odot$\,yr$^{-1}$, when comparing with the population of HMXBs and their associated luminosity.

The galaxy hosts over 50 stellar clusters younger than a few Myr, with masses exceeding $10^5$\,M$_\odot$. 
At its metallicity of $12+\log(\element{O}/\element{H})\approx7.9$ \citep[12\% solar;][]{Metallicity_1985} no other nearby galaxy contains such a large number of massive clusters.
Similar cluster populations are found only in metal-rich mergers (e.g., the Antennae) or more distant galaxies (e.g., Haro 11).
Some clusters exhibit prominent Wolf-Rayet (WR) spectral features.
The brightest, both among the WR clusters and in the galaxy overall, is Cluster 23 (CL23) -- one of the most massive known young clusters with $M>10^7$\,M$_\odot$ and an age of $4$\,Myr \citep{Ostlin_2007}.
It is surrounded by an open superbubble detected in \element{H}{$\alpha$} and has been associated with a large-scale outflow that injects material into the circumgalactic medium.
Furthermore, ESO 338-4 is a candidate Lyman continuum (LyC) galaxy \citep{Leitet2013}. 
Its total \ion{He}{ii}\,$\lambda 4686$\,\AA~line luminosity has been reported as $L^{\text{obs}}_{\ion{He}{ii}~4686,\text{total}}\approx2\times 10^{39}\,\text{erg}~\text{s}^{-1}$ \citep{LidaPap2019}.
The average $\ion{He}{ii}\,\lambda 4686\,\AA/\element{H}\beta$ ratio, as calculated in an aperture of a $4\arcsec$ radius from CL23, is $ 0.019\pm 0.001$.
The galaxy is confidently in the star-forming region of BPT diagrams as presented in the work of \cite{Bik2018}.
In X-rays, three ultra-luminous X-ray sources (ULX) and bright diffuse plasma emission surrounding the galaxy have initially been detected by \cite{LidaPap2019}.

Deep X-ray observations of nearby low-metallicity starbursts are essential to constrain their X-ray binary populations and to search for diffuse hot gas, both of which trace stellar feedback and its impact on the surrounding medium.
In this work, we analyze new deep \textit{Chandra} and \textit{XMM-Newton} observations of ESO 338-4.
This paper is organized as follows.
In Sect.~\ref{sec:Methods}, we describe our observations, the data reduction, and the spectral analysis.
Sect.~\ref{sec:results} presents our results and discusses them in the context of current X-ray luminosity functions, gaseous halos in dwarf starbursts, and the connection between HMXBs and nebular emission.
Sect.~\ref{sec:Conclusions} summarizes our conclusions.
Unless otherwise stated, all X-ray luminosities are reported in the $0.5$--$8$\,keV band.

\section{X-ray observations and data analysis}\label{sec:Methods}

\subsection{Chandra}\label{sec:chandra}

\begin{figure}[h!]
        \centering
        \includegraphics[trim=0 52 280 150, clip,width=\hsize]{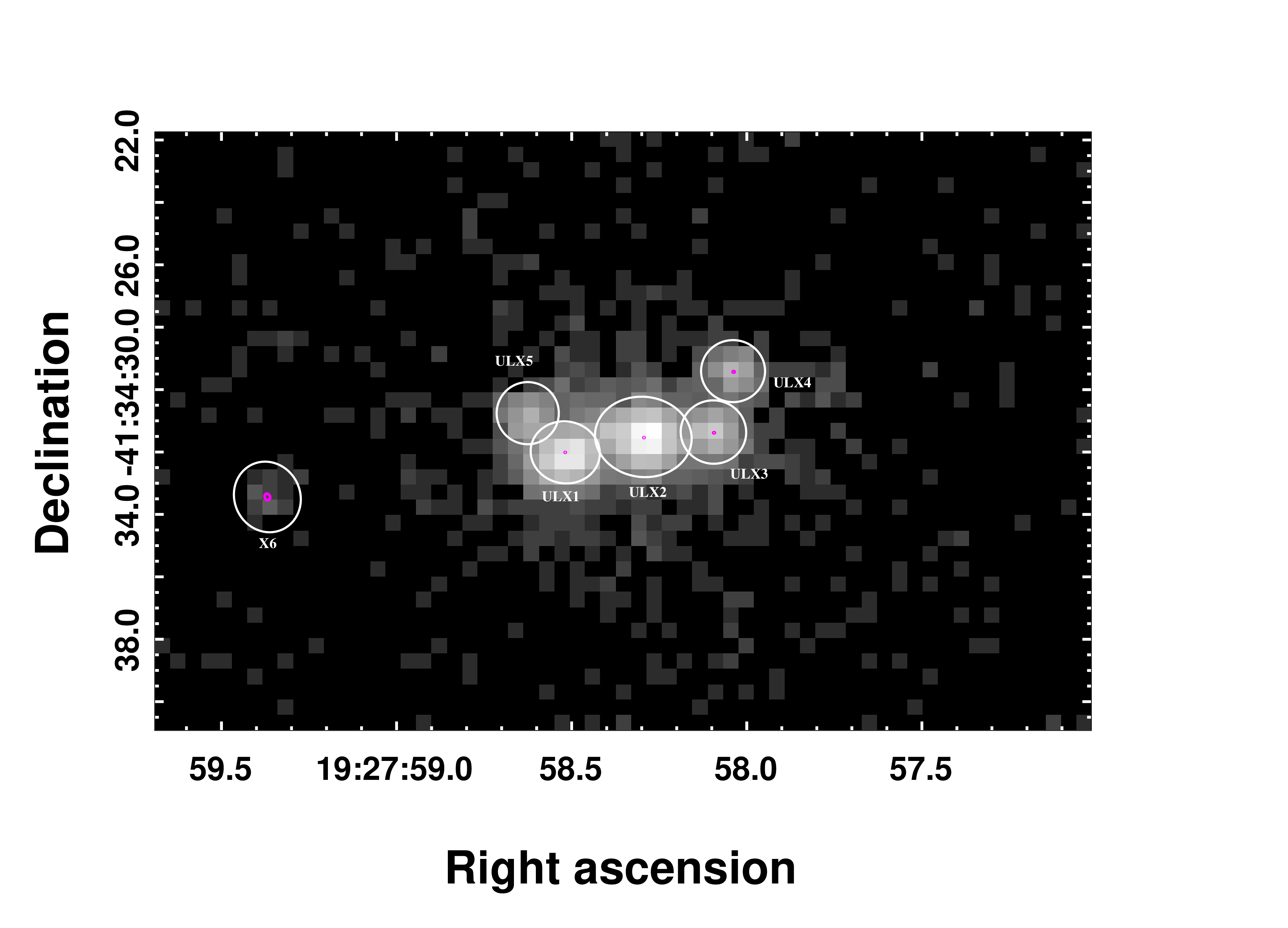}
        \caption{ 
        Merged intensity image of the 16 \textit{Chandra} observations analyzed in this work (the log of the observations is summarized in Appendix~\ref{app:log}).
        The observations have been taken with the ACIS-I array, and the image is exposure corrected.
        We highlight the boundaries of the count distribution of each source at a 3$\sigma$ level through white ellipses.
        Magenta ellipses present the source locations along their 1$\sigma$ uncertainty in position.
        ULX1-4 and X6 were detected by \textit{wavdetect} while ULX5 was manually added as described in the text.
   }
        \label{fig:Chandra_img}
    \end{figure}

\textit{Chandra} observed ESO~338-4 in 2023 16 times (PI: L. Oskinova; Proposal 23620207. Appendix~\ref{app:log} for log of observations) with the ACIS-I array for a combined exposure of 295.35~ks. 
The data were reprocessed applying the latest calibrations with version 4.16 of the Chandra Interactive Analysis of Observations (CIAO)\footnote{\url{https://cxc.cfa.harvard.edu/ciao/}} data analysis software. 
Taking advantage of these deep exposures, we astrometrically
aligned the individual exposures with one another and subsequently with the Gaia EDR3 catalog \citep{Gaia_mand16,Gaia_21,Gaia_mand23,Gaia_23} using the CIAO analysis tools \textit{wcs\_match} and \textit{wcs\_update}.

The final merged and exposure corrected \textit{Chandra} image of ESO~338-4 in the $0.5\text{--}7$\,keV range is presented in Fig.~\ref{fig:Chandra_img}.
To identify X-ray emitters within the galaxy, we used the CIAO source detection algorithm \textit{wavdetect}.
This tool accounts for the instrument's point spread function (PSF) and allows the disentanglement of point sources in close proximity \citep{wavdetect}. 
The resulting count distribution boundaries at a 3$\sigma$ level of each source and its location are indicated in Fig.~\ref{fig:Chandra_img}.
We compute the limiting sensitivity using the CIAO \textit{lim\_sens} function, which returns results in units of photons~cm$^{-2}$~s$^{-1}$. 
To convert this to a flux, we assume an energy-independent photon 
spectrum over the $0.5$--$7$\,keV band.
At the distance of ESO 338-4 ($D=40$\,Mpc), this corresponds to a luminosity  of  $L^{0.5-7~\text{keV}}_{\rm lim}\approx 7 \times 10^{38}\,\text{erg}~\text{s}^{-1}$. 
This value represents the flux significance threshold ($\sim3\sigma$), and is therefore somewhat higher than the formal source detection limit.
Going from east to west, we label the four sources detected by \textit{wavdetect} as ULX 1--4. 
As discussed in Sect.~\ref{sec:spec_anal}, all sources are above the ULX luminosity classification limit of $L_X>10^{39}\,\text{erg s}^{-1}$ \citep[][and references therein]{Kareet2017,King2023}, which corresponds to the critical Eddington luminosity of a $10$~M$_\sun$ black hole. Therefore, for consistency reasons, we refer to them throughout this manuscript as ULXs.

In addition to the four point sources detected by \textit{wavdetect}, we manually identified a fifth source at the easternmost part of the galaxy. 
This detection is motivated by (a) visual inspection of the image and (b) an examination of a diffuse image map of the galaxy. 
For the latter, we follow standard procedures within CIAO.
Thus, we remove the point source contribution from the images and replace the missing pixel values with interpolated values from the local (diffuse) background.
This reveals a bright locus around the proposed location of ULX5. 

Furthermore, \textit{wavdetect} reports a faint sixth source within the areal extent of the galaxy, offset further to the east.
We label it X6, as its luminosity is below the ULX threshold.
With only $\sim20$ counts detected in the merged data set, all below $3$~keV, no spectral or temporal analysis is feasible. 
We characterize it as a soft source and its marginal contribution to the overall count rate has been removed whenever relevant. We do not discuss this source further.
The positions of all six sources are listed in Table~\ref{tab:ULX_prop}.

Three of the \textit{wavdetect} ULXs (ULX1--3) were previously reported by \cite{LidaPap2019}, while our much deeper Chandra exposures ($295.35$\,ks vs. $3.05$\,ks) enable the detection of an additional \textit{wavdetect} ULX (ULX4) and faint X-ray source (X6), as well as the manually identified ULX5.

\begin{table}[h!]
\caption{Locations of X-ray point-sources in ESO 338-4.}                 
\label{tab:ULX_prop}    
\centering                        
\begin{tabular}{c c c }      
\hline\hline              
ESO 338-4 & RA (J2000) & DEC (J2000)  \\         
\hline                      
   ULX1 & $291.993820$ & $-41.575565$ \\    
   ULX2 &$291.992890$& $-41.575428$  \\
   ULX3 &$291.992056$ & $-41.575384$\\
   ULX4 &$291.991828$ & $-41.574842$\\
   ULX5 &$291.994272$ &$-41.575209$ \\
   X6 & $291.99736$ & $-41.57595$\\
\hline                                 
\end{tabular}
\tablefoot{ The coordinates for ULX1--4 and X6 are returned by the \textit{wavdetect} source detection tool. The coordinates of ULX5 correspond to the center of the manually added circular point source region.}
\end{table}

\subsection{XMM-Newton}
\begin{figure}[h!]
\centering
\includegraphics[trim=42 100 150 170, clip,width=\hsize]{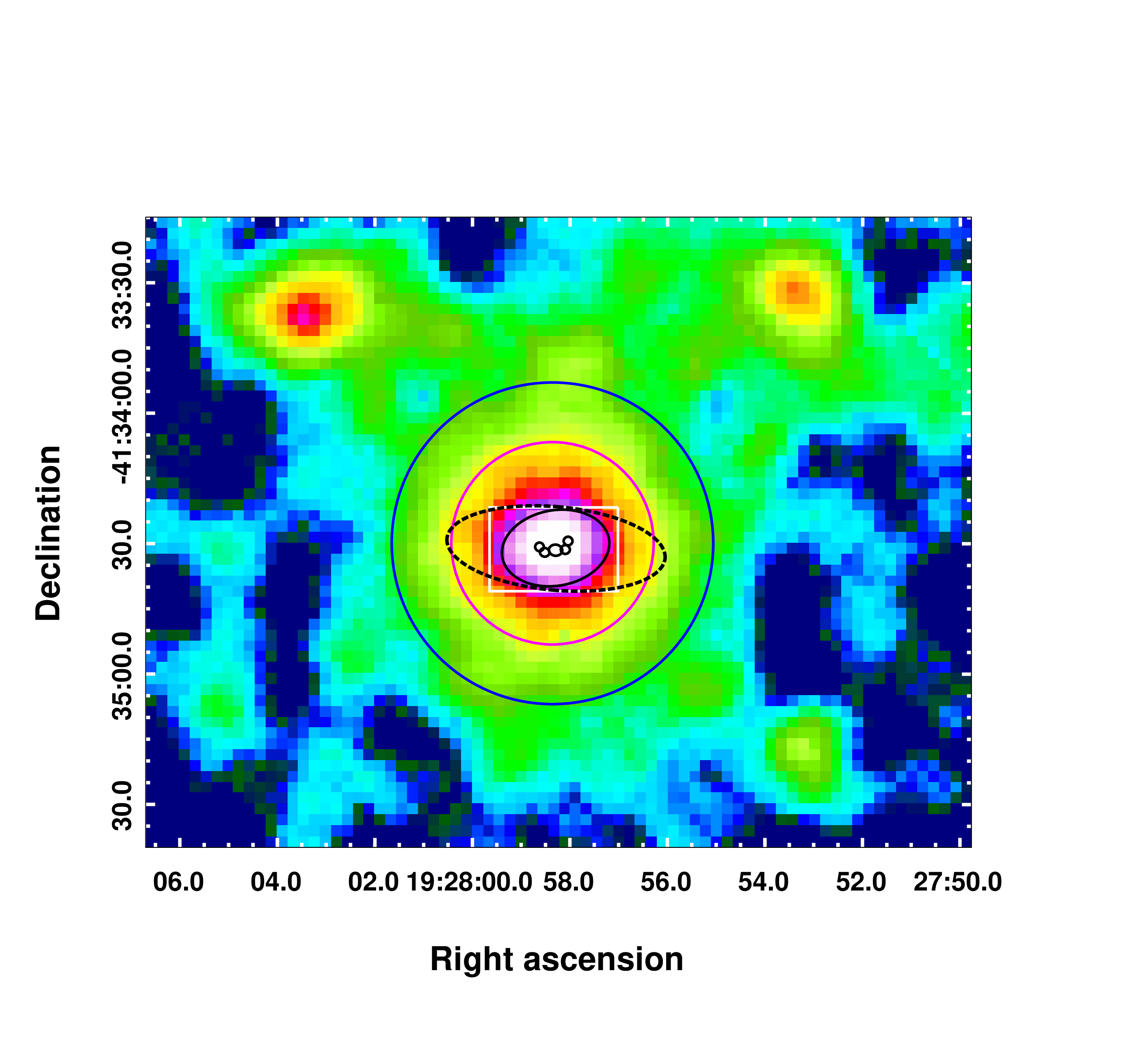}
  \caption{
  Intensity image of the combined \textit{XMM-Newton} EPIC pn and MOS1/2 exposures in the $0.2$--$12$~keV band. 
  The image has been adaptively smoothed and background corrected following standard SAS procedures and is presented with the \textit{gist\_ncar} colormap.
  We highlight with black ellipses the \textit{Chandra} 3$\sigma$ contours presented in Fig.~\ref{fig:Chandra_img}.
  Since \textit{XMM-Newton} does not resolve the individual sources, we depict them through a blended "central region" represented by the larger black ellipse.
  To highlight the different scale between the \textit{Chandra} and \textit{XMM-Newton} images, we note with a white box the extent of Fig.~\ref{fig:Chandra_img}.
  For purposes of spectral fitting in Sect.~\ref{sec:spec_anal} we describe the "galactic halo" by the region between the black ellipse and the magenta circle. 
  Beyond the halo, soft, diffuse structures are visible, extending significantly beyond the extent of the galaxy.
  A second "extended halo" is defined by the region between the magenta and blue circles. Its properties are in agreement with the "galactic halo" and are discussed in Appendix~\ref{app:extended_halo}. 
  The D25 ellipse \citep[HyperLEDA;][]{HyperLedaDistance2014} is noted with dashed black to facilitate comparison with the optical extent of the galaxy.
  The two bright loci in the NW and NE of the galaxy are background sources not associated with ESO 338-4.
 }
    \label{fig:XMM_img}
\end{figure}

In addition to our \textit{Chandra} observations, we take advantage of \textit{XMM-Newton} data using the latter's higher sensitivity at energies below $1$\,keV to study the sources' soft X-ray range and the galactic halo's diffuse emission. 
Besides the 2016 data presented in \cite{LidaPap2019}, \textit{XMM-Newton} observed ESO 338-4 on 17-10-2021 (PI: L. Oskinova; ObsID 0892410101. Appendix~\ref{app:log} for log of observations). 
We use version 21.0.0 of the Science Analysis System (SAS)\footnote{\url{https://www.cosmos.esa.int/web/xmm-newton/sas}} to reprocess and filter our data for Good Time Intervals (GTI). 
Rejecting periods of elevated background, the final useful exposures are $59.58$\,ks for the EPIC pn and $144.48$\,ks for the two combined EPIC MOS cameras. 

Following standard data analysis steps, we background-correct and combine the exposures from the EPIC-pn and the two EPIC-MOS cameras in the $0.4$--$12$\,keV range.
The resulting adaptively smoothed image is presented in Fig.~\ref{fig:XMM_img}. 

The lower angular resolution of the \textit{XMM-Newton} PSF Half Energy Width (HEW of $15$\arcsec and $14$\arcsec for EPIC pn and MOS, respectively) compared to \textit{Chandra} ($0.6$\arcsec) blends the five ULX detected by \textit{Chandra} into an unresolved central region.
This “Central Region” (Fig.~\ref{fig:XMM_img}) was therefore defined to include all five \textit{Chandra}-resolved ULXs within an aperture that exceeds the \textit{XMM-Newton} PSF full width at half maximum of each EPIC camera and is comparable to their respective HEWs.

The sensitivity of \textit{XMM-Newton} allows us to detect soft extended X-ray emission surrounding ESO 338-4 (Fig.~\ref{fig:XMM_img}). 
This hot diffuse gas is referred to as the "galactic halo" throughout this paper.
For the spectral analysis, the halo spectra are extracted from the brightest innermost region between the extent of the galaxy and the boundary presented in Fig.~\ref{fig:XMM_img}. 
We also define a second "extended halo" which extends to larger radii. 
Its spectral properties are consistent with those of the galactic halo and are discussed in Appendix~\ref{app:extended_halo}.

\subsection{Timing analysis}\label{sec:time_anal}

\begin{figure}[h!]
\centering
\includegraphics[width=\hsize]{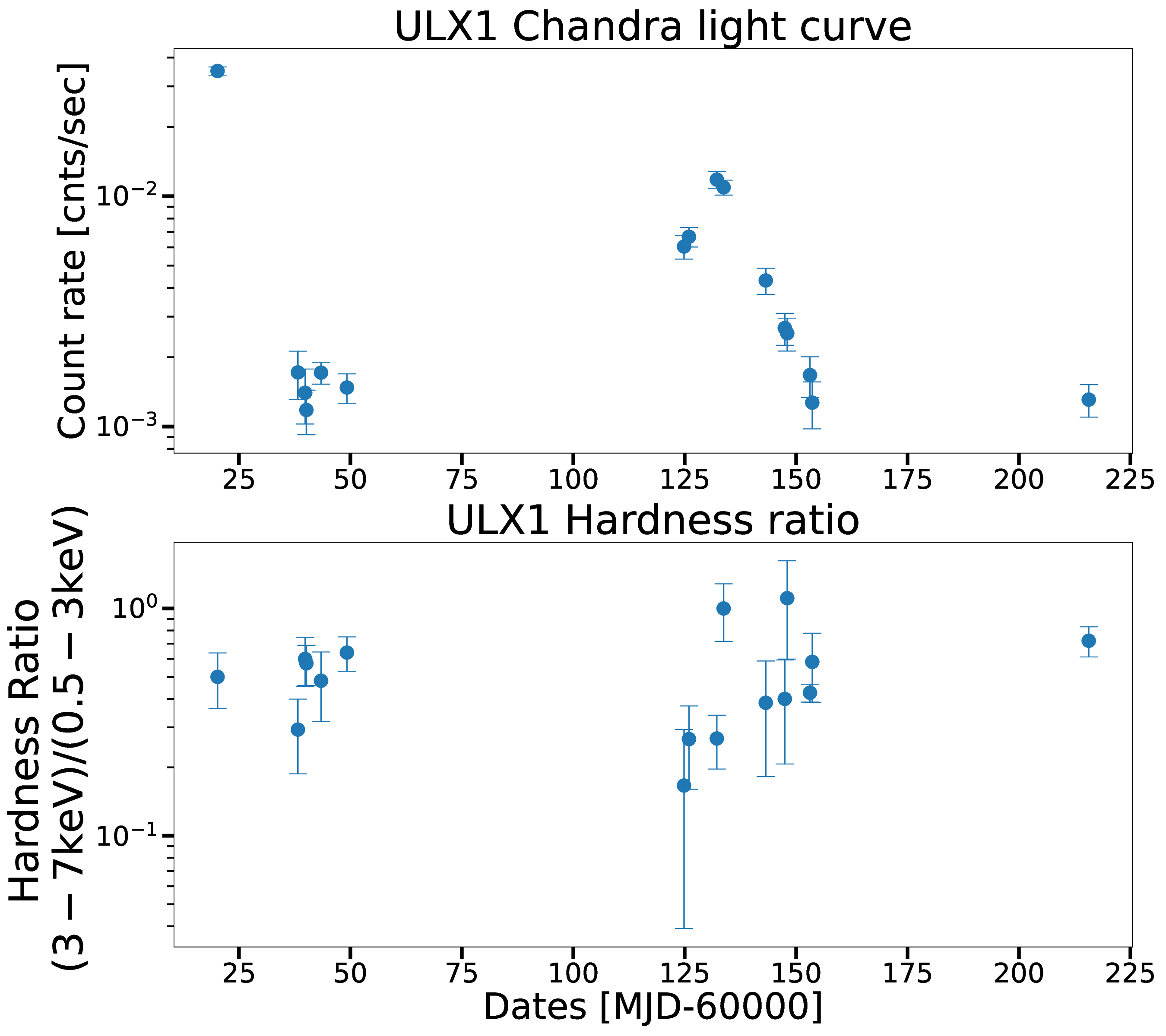}
  \caption{Upper panel: Light curve of ULX1 from the 16 \textit{Chandra} observations obtained in 2023. The curve is constructed from the total count rate in the $0.5$--$7$\,keV range. Lower panel: Hardness ratio of each \textit{Chandra} observation between the $0.5$--$3$\,keV and the $3$--$7$\,keV range. }
    \label{fig:ULX1_var}
\end{figure}

Our 16 \textit{Chandra} observations have been taken throughout approximately 200 days in 2023 (Appendix~\ref{app:log}).
This allows us to search for time variability in individual sources. 
We contrasted the total count rate between observations of each source in the $0.5$--$7$\,keV band.
The X-ray light curves of ULX2--5 (Appendix~\ref{app:var}) do not show signs of strong variability. 
A robust z-score analysis of all observations reveals that ULX3 and ULX4 exhibit no data points deviating beyond 3$\sigma$, while ULX2 and ULX5 each show only a single $>3\sigma$ outlier.
In contrast, ULX1 is variable and displays changes exceeding 1 dex in amplitude (Fig.~\ref{fig:ULX1_var}).
Despite this large variation in brightness, its hardness ratio, defined as $\text{HR}=(3$--$7$\,keV)$/(0.5$--$7$\,keV), stays approximately constant during the same time intervals.
ULX1 dominates the X-ray point source emission in ESO 338-4 during its bright phases and remains the most luminous source when considering count rates averaged over all exposures.

We searched for short-term variability and pulsations using \textit{XMM-Newton} observations.
Lomb–Scargle periodograms were computed from the light curves of the three EPIC cameras in the full band as well as in several soft and hard energy bands, but no statistically significant variability was found.
This indicates that no strong pulsations of ULX1 at a timescale greater than the time resolution of the EPIC cameras in full frame mode are present ($73.4$\,ms for the pn and $2.6$\,s for the MOS cameras).
However, as the \textit{XMM-Newton} observations are not contemporaneous with our \textit{Chandra} data and the point sources are not resolved by \textit{XMM-Newton}, no conclusive argument can be made about the state of ULX1 during them. 
As such, if ULX1 were to be in a low state, pulsations would not be detected, as the other point sources in the unresolved region would dominate the observed X-ray emission.  
Thus, we cannot rule out the possibility that ULX1 is an X-ray pulsar.

\subsection{Spectral analysis}\label{sec:spec_anal}

The next step is to conduct a spectral analysis of the five point sources and the diffuse emission of the galaxy and its halo with the X-ray spectral fitting software \texttt{XSPEC} \citep{xspec1996}. 
The \textit{Chandra} data are fitted in the $0.5$--$7$\,keV range while the \textit{XMM-Newton} observations are fitted in the $0.2$--$10$\,keV range.
We summarize the final best-fit parameters of our analysis in Table~\ref{tab:Spec_pam}, their fit statistic and goodness of fit in Table~\ref{tab:Spec_stat}, and report the X-ray luminosities of each source in Table~\ref{tab:Spec_lum}.
The unfolded \textit{XMM-Newton} spectra for the best-fit models in the central and halo region are presented in Fig.~\ref{fig:XMM_spec_center} and Fig.~\ref{fig:XMM_spec_halo} respectively.

\begin{table}[h!]
\caption{Spectral model parameters for the 5 point sources and the diffuse emission in ESO 338-4.}                 
\label{tab:Spec_pam}    
\centering                       
\begin{tabular}{l  c}     
\hline\hline               
Model Parameter & Best-fit value  \\        
\hline
\noalign{\smallskip}
\multicolumn{2}{c}{Point sources}\\
\noalign{\smallskip}
\hline
\noalign{\smallskip}
\multicolumn{2}{l}{ULX1: \texttt{tbabs$_{\texttt{Gal}}$} $\times$ \texttt{tbabs}$_\text{int}~\times$ (\texttt{diskbb}$_{1} ~+~$\texttt{diskbb}$_2$)}\\
\noalign{\smallskip}
\hline                    
\noalign{\smallskip}
    N$_\text{H, int}$ ($10^{22}$\,cm$^{-2}$)& $0.6\pm0.3$\\
   T$_{\text{in},1}$ (keV)& $1.7\pm0.2$  \\  
   Norm$_1$ & $(8\pm3)\times10^{-4}$\\
   T$_{\text{in},2}$ (keV) & $0.215\pm0.009$ \\  
   Norm$_2$ & $30\pm10$\\

   \hline
\noalign{\smallskip}
\multicolumn{2}{l}{ULX2: \texttt{tbabs$_{\texttt{Gal}}$} $\times$ \texttt{tbabs}$_\text{int}$ $\times$ \texttt{diskpbb}}\\
\noalign{\smallskip}
\hline    
\noalign{\smallskip}
 N$_\text{H, int}$($10^{22}$\,cm$^{-2}$)& $0.1\pm0.2$\\
  T$_{\text{in}}$ (keV)& $1.2\pm0.1$  \\  
   Norm & $(7\pm9)\times10^{-3}$\\
   $\Gamma$ & $1.0\pm0.8$\\
   
\hline
\noalign{\smallskip}
\multicolumn{2}{l}{ULX3: \texttt{tbabs$_{\texttt{Gal}}$}$\times$ \texttt{tbabs}$_\text{int}$ $\times$ \texttt{powerlaw}}\\
\noalign{\smallskip}
\hline 
\noalign{\smallskip}
N$_\text{H, int}$ ($10^{22}$\,cm$^{-2}$)& $0.7\pm0.3$\\
   Norm (photons keV$^{-1}$ cm$^{-2}$ s$^{-1}$ at 1 keV) & $(1.8\pm0.8)\times10^{-5}$  \\
    $\Gamma$  & $2.7\pm0.3$\\
   
\hline
\noalign{\smallskip}
\multicolumn{2}{l}{ULX4: \texttt{tbabs$_{\texttt{Gal}}$}$\times$ \texttt{tbabs}$_\text{int}$ $\times$ \texttt{powerlaw}}\\
\noalign{\smallskip}
\hline 
\noalign{\smallskip}
N$_\text{H, int}$ ($10^{22}$cm$^{-2}$)& $0.9\pm0.6$\\
    Norm (photons keV$^{-1}$ cm$^{-2}$ s$^{-1}$ at 1 keV)& $(1.0\pm0.7)\times10^{-5}$  \\
    $\Gamma$  & $(2.5\pm0.6)$\\
     
\hline
\noalign{\smallskip}
\multicolumn{2}{l}{ULX5: \texttt{tbabs$_{\texttt{Gal}}$}$\times$ \texttt{tbabs}$_\text{int}$  $\times$ \texttt{powerlaw}}\\
\noalign{\smallskip}
\hline
\noalign{\smallskip}
N$_\text{H, int}$ ($10^{22}$cm$^{-2}$)& $2.1\pm0.8$\\
   Norm (photons keV$^{-1}$ cm$^{-2}$ s$^{-1}$ at 1 keV) & $(1.6\pm1.2)\times10^{-6}$  \\
    $\Gamma$  & $2.5 \pm 0.6$ \\
      
   \hline
\noalign{\smallskip}
\multicolumn{2}{c}{Diffuse sources}\\
\noalign{\smallskip}
\hline 
\noalign{\smallskip}
\multicolumn{2}{l}{Halo: \texttt{tbabs$_{\texttt{Gal}}$} $\times$ (\texttt{APEC}$_1$ + \texttt{APEC}$_2$)}\\
\noalign{\smallskip}
\hline
\noalign{\smallskip}
kT$_1$ (keV) & $0.26\pm0.02$\\
Norm$_1$ (cm$^{-3}$)& $(2.7\pm0.4)\times10^{-5}$\\
kT$_2$ (keV) &$3.5\pm0.3$\\
Norm$_2$ (cm$^{-3}$)& $(6.1\pm0.2)\times10^{-5}$\\

\hline
\noalign{\smallskip}
\multicolumn{2}{l}{Central: \texttt{tbabs$_{\texttt{Gal}}$} $\times$ (\texttt{APEC}$ + $ Point-source contribution)}\\
\noalign{\smallskip}
\hline   
\noalign{\smallskip}
kT (keV)& $0.262\pm0.007$ \\
Norm (cm$^{-3}$)& $(3.7\pm0.1)\times10^{-4}$\\

 \hline
\end{tabular}
\tablefoot{The results are derived by fitting the spectra of the \textit{Chandra} and \textit{XMM-Newton} observations of Appendix~\ref{app:log}. The Galactic foreground absorption \texttt{tbabs$_{\texttt{Gal}}$} is fixed at the value of $N_{\element{H}}=5\times 10^{20}\,\text{cm}^{-2}$.}
\end{table}

\begin{table}[h!]
\caption{Goodness-of-fit statistics for the spectral models fitted to the \textit{Chandra} and the \textit{XMM-Newton} observations of ESO~338-4. 
}               
\label{tab:Spec_stat}    
\centering                       
\begin{tabular}{c c c c}      
\hline\hline               
\noalign{\smallskip}
ESO 338-4 & C-statistic & d.o.f. & $p_{\rm null}$  \\        
\noalign{\smallskip}
\hline
\noalign{\smallskip}
\multicolumn{4}{c}{\textit{Chandra}}\\
\noalign{\smallskip}
\hline
\noalign{\smallskip}
   ULX1&$222$ & $207$ & $0.318$\\    
   ULX2&$218$ & $265$ & $0.962$ \\
   ULX3&$79$ & $72$ & $0.083$\\
   ULX4&$43$ & $38$ & $0.448$\\
   ULX5&$46$ & $46$ &$0.679$ \\
\hline
\noalign{\smallskip}
\multicolumn{4}{c}{\textit{XMM-Newton}}\\
\noalign{\smallskip}
\hline
\noalign{\smallskip}
   Central region&$525$ &$422$ & $0.002$\\
   Halo region&$232$ &$213$ & $0.209$ \\
\hline                                  
\end{tabular}
\tablefoot{For each region, we report the C-statistic, the number of degrees of freedom (d.o.f.), and the null-hypothesis probability ($p_{\rm null}$).}
\end{table}

\begin{table}[h!]
\caption{Luminosities of ULX and diffuse emission in ESO 338-4.}                
\label{tab:Spec_lum}   
\centering                        
\begin{tabular}{c c c c}      
\hline\hline              
\noalign{\smallskip}
ESO 338-4 & $\log(L^{\rm obs}_X$) &$\log(L^{\rm cor}_X$) & $\log(L^{\text{model}}_{0.054\text{--}12\text{\,keV}}$)  \\       
\noalign{\smallskip}
&(erg s$^{-1}$)&(erg s$^{-1}$)&(erg s$^{-1}$)\\
\noalign{\smallskip}
\hline                      
   ULX1&$40.47$ & $41.14$ & $41.5$\\    
   ULX2&$40.45$ & $40.49$ & $40.5$ \\
   ULX3&$39.63$ & $40.04$ & --\\
   ULX4&$39.37$ & $39.78$ & --\\
   ULX5&$39.46$ & $40.01$ &-- \\
   Galaxy diffuse&$40.15$ &$40.26$& $40.7$\\
   Galaxy total& $40.91$ &$41.33$ & $41.6$\\
   Halo& $40.13$ &$40.17$ & -- \\
\hline                                  
\end{tabular}
\tablefoot{We report both the observed luminosities $L^{\rm obs}_X$, uncorrected for absorption, and the absorption-corrected values $L^{\rm cor}_X$ .
The $0.5\text{--}8$\,keV X-ray luminosities ($L_X$) are calculated from the best-fit spectral models applied to the data (Table~\ref{tab:Spec_pam}).
The luminosities in the $0.054\text{--}12$\,keV range are derived by extrapolating the absorption-corrected models.
The galaxy's $L^{\text{model}}_{0.054\text{--}12\text{\,keV}}$ is corrected for the intrinsic absorption in case of ULX1 and 2 but not for ULX3, 4, and 5, as extending a power law to UV energies would yield unrealistically high luminosities.}
\end{table}

\subsubsection{Spectral analysis -- \textit{Chandra}}

During the spectral analysis, we devote special attention to the background treatment.
First, a source-free region on the same CCD as the target galaxy is identified.
The instrumental background extracted from this region is subtracted from all spectra.
Next, we address the local background of each point source, treating it as a local diffuse emission filling the galaxy.
These regions are defined as annuli around each source region with inner and outer radii of 0.5 and 2.5 pixels\footnote{2 pixels for ULX5 to avoid including local background from ULX1 and ULX2.} from the source region boundary, where contributions from neighboring point sources have been excluded. 

Using CIAO, we separately co-add the 16 source spectra and local background spectra for each ULX.
Following our timing analysis (Sect.~\ref{sec:time_anal}), we analyze an average state of ULX1.

So far, there are no universally accepted physically motivated spectral models of ULXs.
The complexity of models ranges from simple power laws to sophisticated combined models requiring multiwavelength data.
Furthermore, the assumptions on spectral shape at energies below those observed by the X-ray telescopes (i.e., at energies lower than $0.2$\,keV) crucially affect the estimates on the number of \ion{He}{II} ionizing photons. 
In our spectral modeling, we considered the following three spectral models (their \texttt{XSPEC} names in parentheses): a simple power law (\texttt{powerlaw}), 2 standard multicolored black-body disks (\texttt{diskbb}+\texttt{diskbb}), and multicolor black-body disks with variable power-law dependence for disk temperature, $T(r)$ (\texttt{diskpbb}). For the local background emission, we consider collisionally ionized plasma (\texttt{APEC}), where we set the metallicity to $0.12\,Z_\odot$ for ESO 338-4 \citep{Metallicity_1985}.
Lastly, we considered absorption components (\texttt{tbabs}) for the Galactic foreground \citep[$N_H=5\times\,10^{20}$cm$^{-2}$;][]{nh_eso2005}, and a freely varying internal absorption within the galaxy.

We simultaneously model the point sources and their local background in the \textit{Chandra} observations, grouping the data to a minimum of 5 counts per bin, and using C-statistics \citep{Cash1979}. The \textit{Chandra} spectra are presented in Appendix~\ref{app:CH_spec}. 
All spectral models of the local background feature a hot plasma emission with a temperature of $2$--$3$\,keV. 
We find that the \texttt{diskpbb} model provides the best fit for ULX2, the second brightest point source in ESO 338-4. Although some parameters are not well-constrained, caused by \textit{Chandra's} limited sensitivity below $1$\,keV, the other models provide a worse fit. 
For ULX3--5, insufficient counts prevent a characterization of the spectral shape, and we therefore adopt the simple absorbed power-law model.
All three spectral models feature equally agreeable goodness-of-fit statistics to ULX1.
However, similar to ULX2, the best-fit parameters are not well constrained. 
Furthermore, the hardness ratio between the $0.5$--$3$\,keV and the $3$--$7$\,keV range (Fig.~\ref{fig:ULX1_var}) is not correlated with the count rate variability in a way that would favor any specific model interpretation. 
Thus, to choose a model for ULX1, we use the \textit{XMM-Newton} data (Sec.~\ref{sec:spec_anal_XMM}) as it provides spectral information in soft X-rays.
We assume that ULX1 dominates the count rate among point sources in the extraction region, allowing its spectrum to be modeled using these data.

\subsubsection{Spectral analysis -- \textit{XMM-Newton}}\label{sec:spec_anal_XMM}
To analyze the diffuse emission in the galactic halo, we adopt a two-component plasma model (\texttt{APEC}+\texttt{APEC}). This choice is motivated by the presence of both hard and soft plasma components within the galaxy. The hard component is detected as faint local background in the \textit{Chandra} observations but is unresolved in the \textit{XMM-Newton} data. The soft component is revealed by \textit{XMM-Newton}, thanks to its higher sensitivity below $1$\,keV compared to \textit{Chandra}.
For both components, the host galaxy's metallicity of $Z=0.12\,Z_\odot$ is used.
The final best-fit model for the halo is \texttt{tbabs}${_\text{Gal}}$ $ \times$ (\texttt{APEC}$_1$ $+$ \texttt{APEC}$_2$) with the corresponding unfolded spectrum from \textit{XMM-Newton} shown in Fig.~\ref{fig:XMM_spec_halo}.

The limited angular resolution of \textit{XMM-Newton} prevents spatial separation of point sources from the surrounding diffuse emission in ESO 338-4.
Consequently, we fit the blended central region of our \textit{XMM-Newton} 2021 observations (Fig.~\ref{fig:XMM_img}) with the combination of the individual ULX models and an \texttt{APEC} component to account for unresolved hot gas within the galaxy, not detectable with \textit{Chandra}. 
We refrain from adding a second \texttt{APEC} component, as in the halo analysis, since the corresponding hard thermal contribution is negligible with respect to the dominant ULX emission (Appendix~\ref{app:CH_spec}) and would therefore not yield well-constrained parameters.

\begin{figure}[h!]
\centering
\includegraphics[trim=45 25 25 142, clip,angle=-90,width=\hsize]{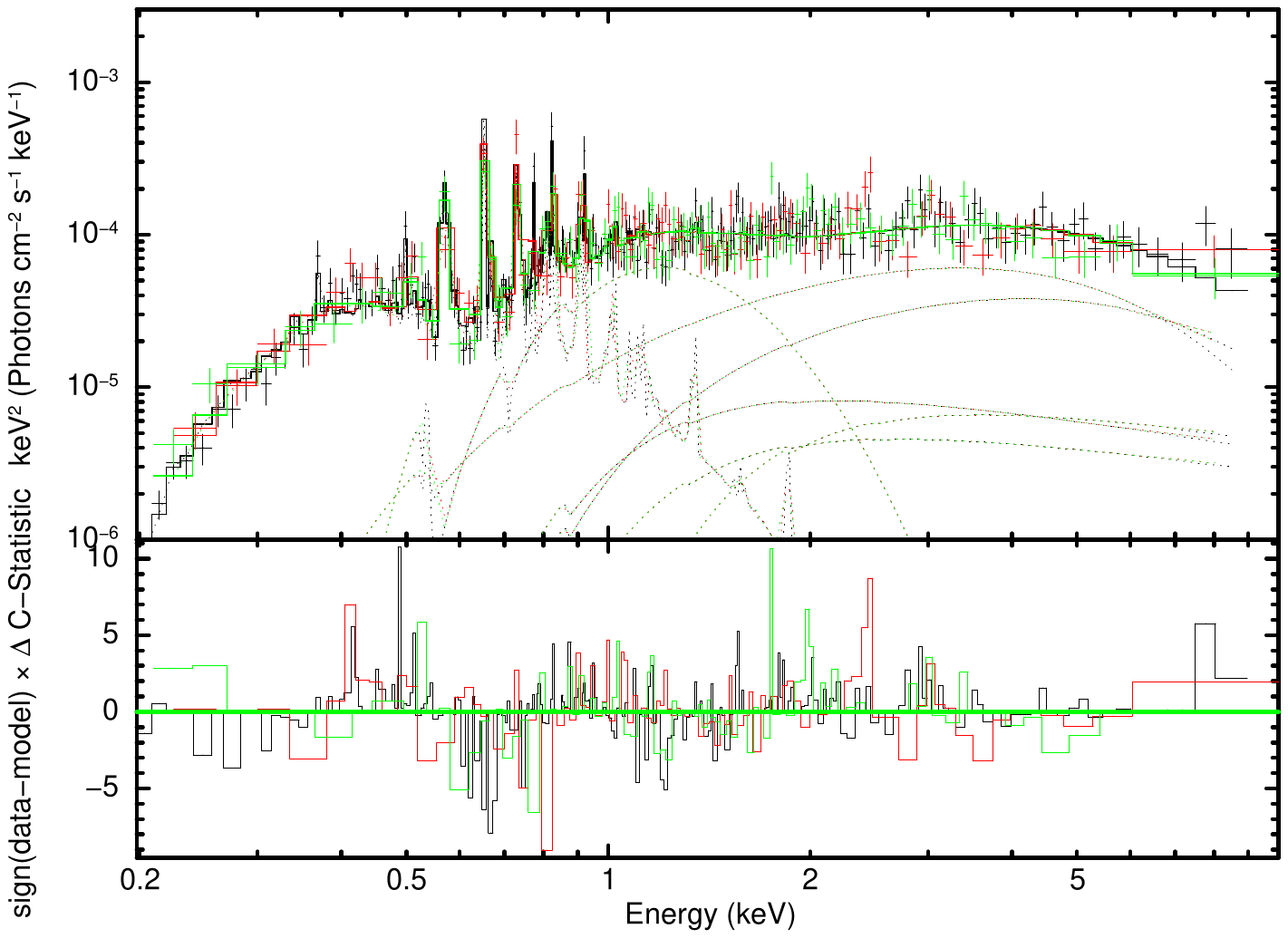}
\begin{picture}(0,0)
    \put(-65,175){\makebox(0,0){Central Region}} 
\end{picture}
  \caption{Upper panel: \texttt{XMM-Newton} EPIC pn, MOS1, and MOS2 spectra of ESO 338-4 (black, red, and green curves, respectively) with 3$\sigma$ errors. The best-fit model (described in Sect.~\ref{sec:spec_anal_XMM}) of the galaxy is shown with solid lines. The individual best-fit model components are shown with dashed lines. The model parameters are given in Table~\ref{tab:Spec_pam}. Lower panel: residuals between the data and the best-fit model. }
    \label{fig:XMM_spec_center}
\end{figure}

\begin{figure}[h!]
\centering
\includegraphics[trim=45 25 25 142, clip,angle=-90,width=\hsize]{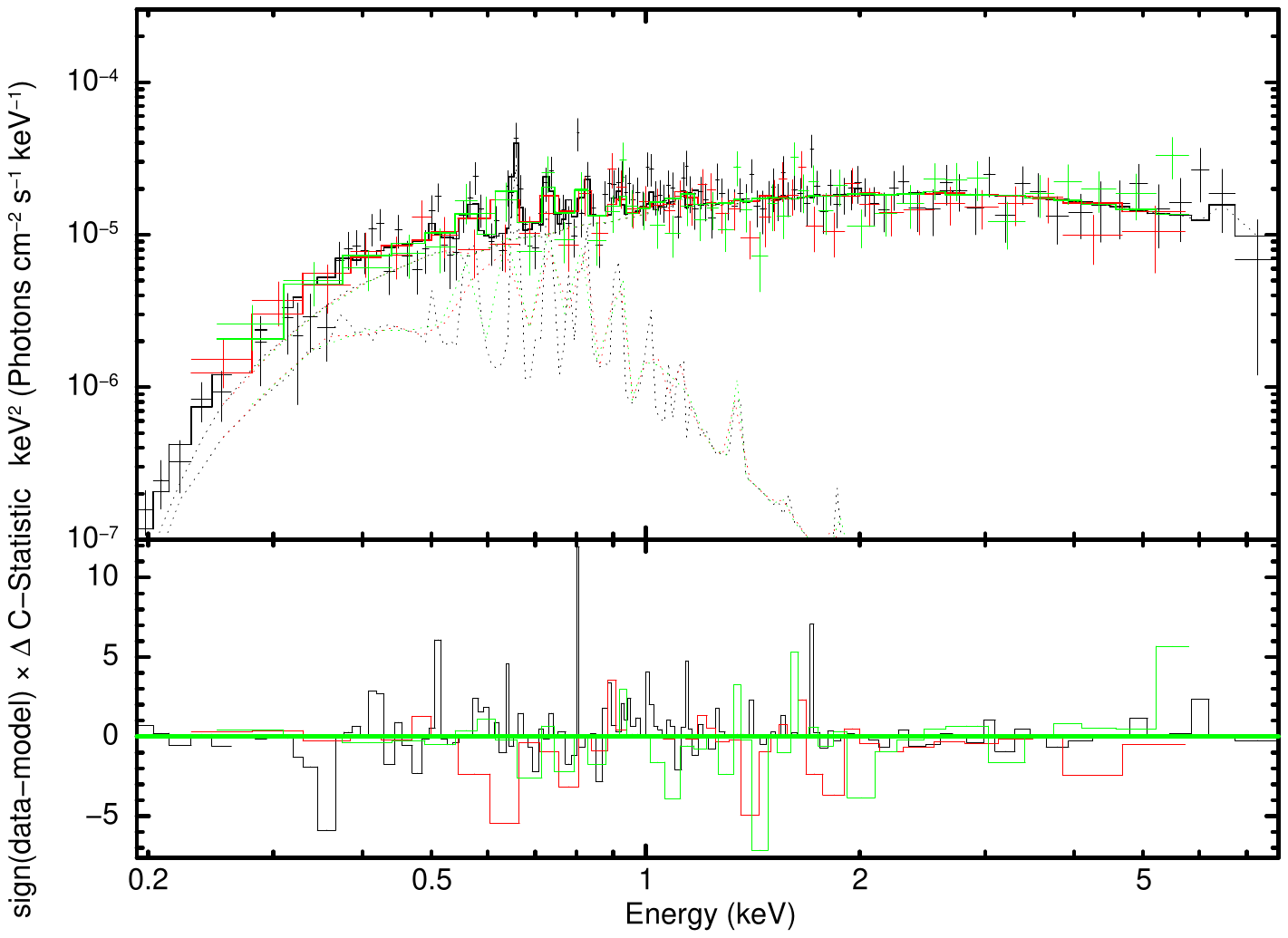}
\begin{picture}(0,0)
    \put(-65,175){\makebox(0,0){Halo Region}} 
\end{picture}
  \caption{Same as Fig.~\ref{fig:XMM_spec_center} for the galactic halo of ESO 338-4 defined in Fig.~\ref{fig:XMM_img}. The best-fit model is described in Sect.~\ref{sec:spec_anal_XMM} and the model parameters are given in Table~\ref{tab:Spec_pam}. }
    \label{fig:XMM_spec_halo}
\end{figure}

We used Cash statistics and a minimum of 15 counts per bin.
We note that, due to the longer GTI of the 2021 observations compared to those of 2016, we focus our analysis solely on the former. 
However, all subsequent results are confirmed by simultaneously fitting the 2016 along the 2021 data. 
During the fitting procedure, we freeze the parameters of ULX2--5\footnote{Due to its lower luminosity compared to ULX1, unfreezing model parameters for ULX2 does not yield meaningful constraints.} and the constrained parameters of ULX1.
All three spectral models for ULX1 (\texttt{powerlaw, diskpbb, diskbb+diskbb}) feature comparable fit statistics and are well-constrained. 
However, among those, only the \texttt{diskbb}$+$\texttt{diskbb} model yields a hot gas temperature for the \texttt{APEC} component that is consistent with the soft thermal emission detected in the halo. 
We therefore adopt \texttt{diskbb}$+$\texttt{diskbb} as the preferred spectral model for ULX1. 
The final best-fit model for the central region is \texttt{tbabs}${_\text{Gal}}$ $\times$ (\texttt{tbabs}$_{\text{ULX1}}$ $\times$ (\texttt{diskbb}$_1~+$ \texttt{diskbb}$_1$) + \texttt{tbabs}$_{\text{ULX2}}$ $\times$ \texttt{diskpbb} $+$  \texttt{tbabs}$_{\text{ULX3}}$ $\times$ \texttt{powerlaw}$_1$ $+$ \texttt{tbabs}$_{\text{ULX4}}$ $\times$ \texttt{powerlaw}$_2$ $+$ \texttt{tbabs}$_{\text{ULX5}}$ $\times$ \texttt{powerlaw}$_3$ $+$ \texttt{APEC}) with the corresponding unfolded spectrum from \textit{XMM-Newton} shown in Fig.~\ref{fig:XMM_spec_center}.

It is common for halo gas to exhibit lower metallicities than the galaxy’s ISM \citep{tuli2017}. To assess the feasibility of such a scenario, we explored models with metallicities in the range $Z=0.02\text{--}0.12\,\text{Z}_\sun$. This analysis confirmed that our adopted model provides the most robust description of the data, while lower-metallicity models yield either unconstrained parameters or non–physically motivated solutions.


As discussed in Sect.~\ref{sec:time_anal}, the offsets in observational times and the differences between the angular resolutions of \textit{Chandra} and \textit{XMM-Newton} limit the reliability of using one data set to inform the analysis of the time-variable ULX1 in the other.
However, varying freely all spectral model parameters of ULX1 in the \textit{XMM-Newton} fit, returns a hard \texttt{diskbb} component with comparable parameters to those derived from the \textit{Chandra} data.
This consistency supports the interpretation that the model captures the average spectral properties of the source.

The spectral models described above allow us to predict the luminosity between the ionization potential of \ion{He}{ii} of $54$\,eV ($4$\,Ry) and the upper range of \textit{XMM-Newton}, $12$\,keV (Table~\ref{tab:Spec_lum}).
We find the total $0.054$--$12$\,keV luminosity for ESO 338-4 is $\log(L^{\text{model}}_{0.054\text{--}12\text{\,keV}}$[erg s$^{-1}])=41.6$, with ULX1 contributing $\log(L^{\text{model}}_{0.054\text{--}12\text{\,keV}}$[erg s$^{-1}])=41.5$ (80\%; The contribution of ULX1 to the total absorption-corrected X-ray luminosity of the galaxy is 65\%) to it.
We do not report values for ULX3, 4, or 5, as their model description of a power-law would yield unrealistically high luminosities at UV energies.
For this reason, these point sources are also not corrected for intrinsic absorption in the derivation of the galaxy luminosity.

\subsubsection{The $E>1$\,keV halo component}

\begin{figure}[h!]
\centering
\includegraphics[trim=70 45 25 130, clip,angle=-90,width=\hsize]{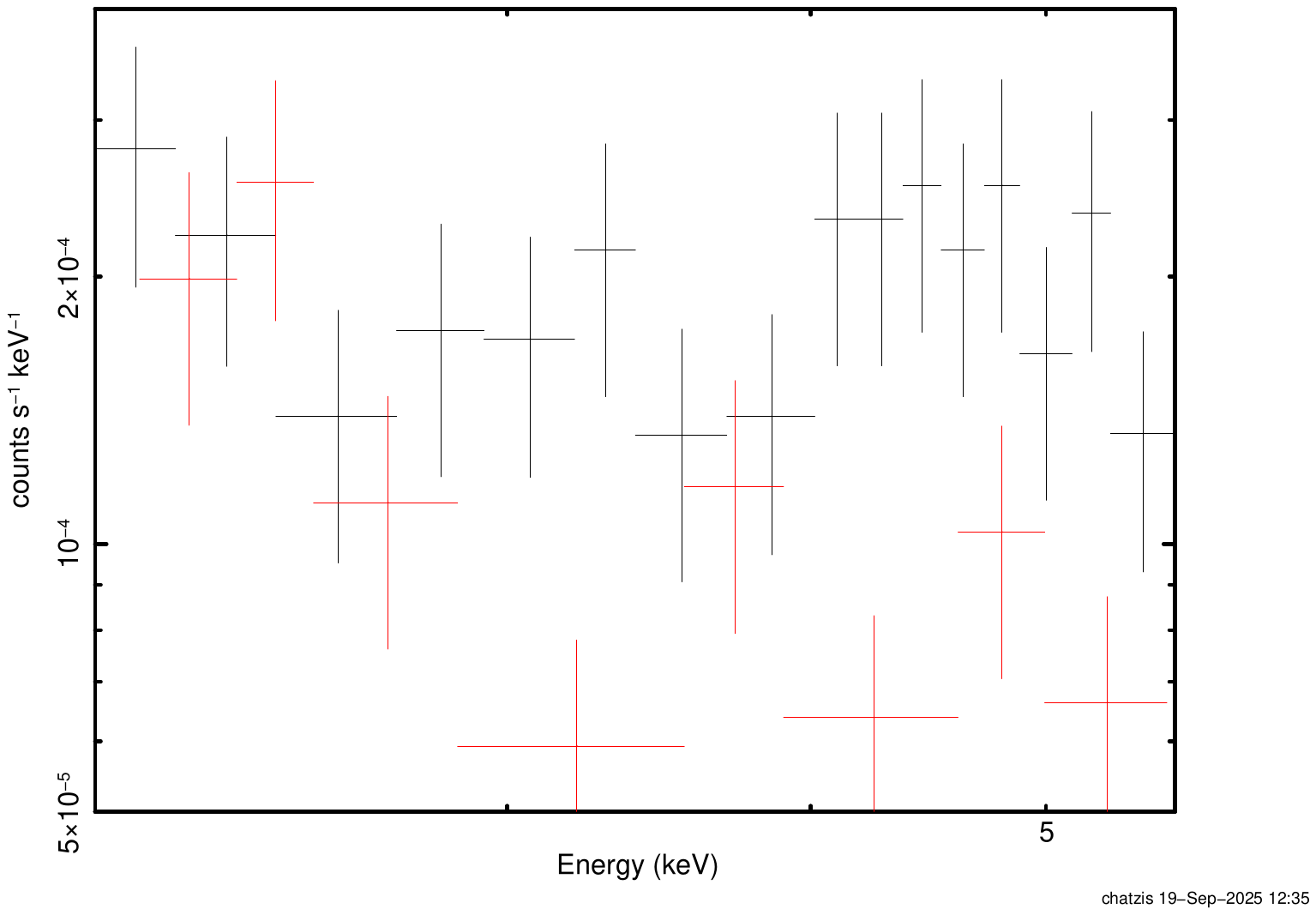}
\caption{
Co-added spectrum of the halo region (black) compared to the corresponding local background (red), extracted from the 16 \textit{Chandra} observations. 
A weak but systematic excess is visible in the $2$--$6$\,keV band, consistent with the hotter ($kT\simeq3.5$\,keV) component inferred from the \textit{XMM-Newton} analysis (Table~\ref{tab:Spec_pam}). 
The spectrum is shown without background subtraction to illustrate the level of the excess relative to the local background.
}
\label{fig:Ch_halo}
\end{figure}

An important caveat concerns the origin of the hotter ($kT\simeq3.5$\,keV) plasma component.
One possible explanation is that it reflects the broad \textit{XMM-Newton} PSF wings, allowing counts from the central region to contribute to the halo spectrum.
While this possibility cannot be excluded without a dedicated PSF simulation, which is beyond the scope of this paper, our data suggest that the component is not entirely instrumental.

We fitted the \textit{XMM-Newton} halo region with two alternative models:  \texttt{tbabs$_{\texttt{Gal}}$} $\times$ (\texttt{APEC}$_1$ + \texttt{APEC}$_2$) and  \texttt{tbabs$_{\texttt{Gal}}$} $\times$ (\texttt{APEC} + \texttt{const} $\times$ (Point-source contribution)).
The parameters and fit statistic of the former are presented in Table~\ref{tab:Spec_pam} and Table~\ref{tab:Spec_stat}. 
The latter represents soft plasma emission with contamination from the ULX population, with parameters fixed according to Table~\ref{tab:Spec_pam}. 
This fit yields \texttt{APEC} parameters of $\text{kT}=(0.244\pm0.008)$\,keV, $\text{Norm}=(7.2\pm0.3) \times 10^{-6}$, and a \texttt{constant} factor of $0.170\pm0.005$; a PSF fraction of roughly $20\%$.
Its C-statistic for this fit is 247 for 214 d.o.f. and $p_{\rm null}=0.062$.
Both models provided comparable fit statistics, indicating that the data do not allow us to distinguish uniquely between unresolved point-source contamination and a hot diffuse plasma.

Additional support for a non-instrumental contribution comes from the co-added spectrum of the 16 \textit{Chandra} observations extracted from the halo region (Fig.~\ref{fig:Ch_halo}). 
Despite \textit{Chandra's} much narrower PSF, the co-added spectrum shows a weak but persistent excess relative to the local background in the $2$--$6\ \mathrm{keV}$ range; the best-fit hot \texttt{APEC} temperature from the \textit{XMM-Newton} fit ($kT \simeq 3.5\ \mathrm{keV}$) lies inside this energy interval, where the \textit{Chandra} halo spectrum and background begin to diverge. The excess is marginal and could be influenced by statistical fluctuations, but its persistence in the co-added \textit{Chandra} data argues against it being purely noise.
We therefore include the hot component in our modeling but regard it as tentative until deeper data or future PSF simulations can clarify its origin.

Previous halo studies have mainly targeted spiral or elliptical galaxies, often limited to soft-band analyses or modeling harder emission as unresolved sources, typically with shorter exposures \citep{Strickland2004a,Strickland2004b,Boroson2011,LiWang2013}.
In these circumstances, a faint, extended hard component would be difficult to detect. The absence of similar reports should therefore be understood as reflecting methodological and sample differences rather than as evidence against a tentative hot component in low-metallicity dwarf starbursts.

\section{Results and discussion}\label{sec:results}

\subsection{Time variability of ULX1}

The long-term variability of ULXs is an open topic and is subject to debate.
We test whether pure changes in the internal absorption column density of the spectral model of ULX1 (Table~\ref{tab:Spec_pam}; Parameters derived in Sect.~\ref{sec:spec_anal}) can explain the observed variability of ULX1. 
To reproduce a 1 dex change in brightness, the internal absorption would have to increase from zero to \texttt{tbabs}$_\text{int} = 3 \times 10^{22}\,\text{cm}^{-2}$.
This would result in a shift from $\text{HR}\approx0.09$ to $\text{HR}\approx2.4$, in contrast to the minimal changes observed in the hardness ratio of ULX1 (Fig.~\ref{fig:ULX1_var}).
If we assume a non-zero initial internal absorption, consistent with the local \ion{H}{i} column density inferred from Ly$\alpha$ measurements, even larger changes in the column density would be required. This would produce correspondingly stronger variations in the hardness ratio, which are not observed.
Therefore, we rule out variable internal absorption as the cause of ULX1's flux variability.

\subsection{HMXB X-ray luminosity in low-Z galaxies}\label{sec:XLF}

Recent studies indicate that the X-ray luminosity of low-metallicity galaxies may be substantially higher than previously estimated \citep[e.g,][]{OLDXFL_fornasini2020,OldXFL_bret2022, Kyritsis2025}.
\cite{Lehmer2024} presented a scaling relation between the integrated X-ray luminosity of high mass X-ray binaries of galaxies, and their SFR and $Z$, defining the $L_X/\text{SFR--}Z$ plane.
By applying absorbed broken power laws to point sources observed with \textit{Chandra}, they reconstruct the XLFs of individual galaxies based on their SFR and $Z$.
ESO 338-4 is included in this sample, albeit with a somewhat higher metallicity of $12+\log(\text{O}/\text{H})=7.99$, and a reported luminosity of $\log(L_X [\text{erg}~\text{s}^{-1}]) = 40.3 \pm 0.2$.

In our analysis of ESO 338-4, we measure from our best-fit model of the \textit{XMM-Newton} data a cumulative X-ray luminosity from point sources of $\log(L_X\,[\text{erg}~\text{s}^{-1}]) = 41.29$.
For comparison, \cite{Lehmer2024} report a luminosity without correction for intrinsic absorption. 
Applying only the Galactic foreground absorption correction to our spectral model yields $\log(L_X\,[\text{erg}~\text{s}^{-1}]) = 40.84$.
The X-ray scaling relations of \cite{Lehmer2024} are calibrated using the average SFRs over the past $100\text{--}125$\,Myr, as derived from SED-based star formation histories.
To enable a direct comparison with their best-fit $L_X/\text{SFR--}Z$ relation, we therefore adopt their inferred value of $\text{SFR}_{125\,\text{Myr}}=0.6\,\text{M}_\odot\,\text{yr}^{-1}$ for ESO 338-4. 
We report $\log(L_X/\text{SFR}\,[\text{erg}~\text{s}^{-1}(\text{M}_\odot\,\text{yr}^{-1})^{-1}])$ ratios of 41.5 and 41.1, respectively.

We derive $L_X/\text{SFR}$ values that lie slightly above the average relation of the fundamental $L_X/\text{SFR--}Z$ plane presented in \cite{Lehmer2024}, but within their $3\sigma$ confidence interval. 
Based on the scaling relation of \cite{King2023}, a galaxy with a SFR equal to the mean SFR of ESO 338-4 in this study ($2~\text{M}_\odot\,\text{yr}^{-1}$) would be expected to host roughly one bright ULX. 
In contrast, ESO 338-4 contains four sources with $\log(L^{\rm cor}_X~[\text{erg~s}^{-1}])>40$, including one exceeding $\log(L^{\rm cor}_X~[\text{erg~s}^{-1}])>41$. 
Nevertheless, this mild excess in luminous sources is consistent with Poisson fluctuations, which are common at the high-$L_X$ end of X-ray luminosity functions. 
Therefore, ESO 338-4 does not appear to be an outlier relative to XLF predictions based on metallicity and star-formation history.

To estimate the contribution from the cosmic X-ray background (CXB), we use the results presented in \cite{Kim2007}. 
They analyze combined data from the Chandra Multiwavelength Project and the Chandra Deep Field-South surveys, reporting CXB flux densities in units of erg~s$^{-1}$ cm$^{-2}$ deg$^{-2}$. 
Multiplying these values by the projected area of ESO 338-4 (indicated by the black ellipse in Fig.~\ref{fig:XMM_img}) and its distance, we find that the CXB luminosity is less than $1$\% of the observed X-ray luminosity, indicating that background corrections do not play a significant role in luminosity derivations.

\subsection{The galactic halo of ESO 338-4}\label{sec:Halo}

\subsubsection{Morphology}
\setlength{\fboxsep}{0pt} 
\setlength{\fboxrule}{1pt} 

\begin{figure}[h!]
    \centering
    \begin{minipage}[c]{\hsize}
    \makebox[0pt][l]{
        \hspace*{0pt} 
        \fbox{\makebox[0.95\hsize][l]{
            \includegraphics[trim=0 0 42 0, width=\hsize]{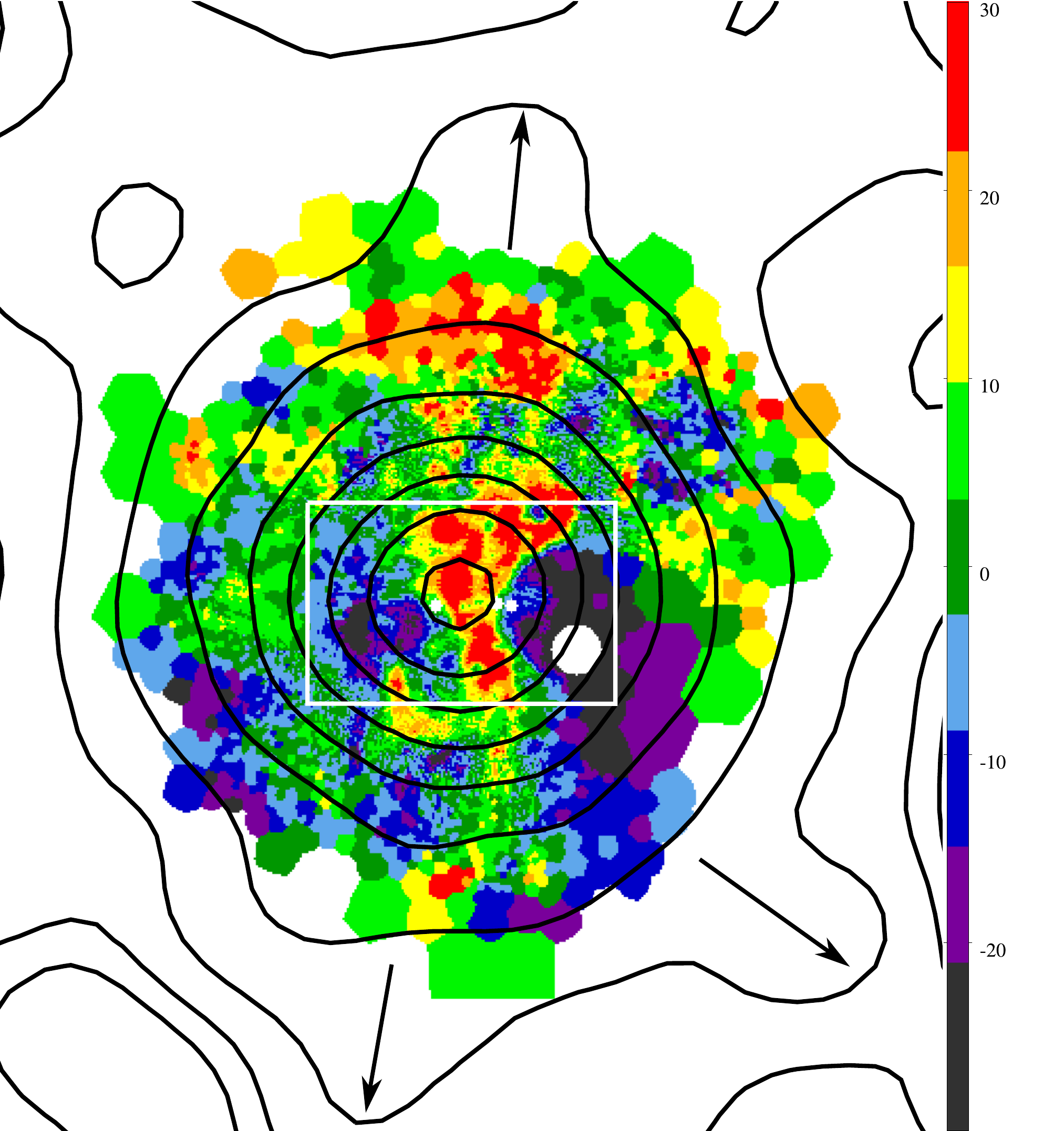}
        }}
    }
    \end{minipage}
    \caption{\element{H}$\alpha$ velocity map of ESO 338-4 presented in \cite{Bik2018}. The colormap indicates the blue and red shifted regions from $-30$ to $30$ km s$^{-1}$.
    CL23 is located at the center of the image, at the base of the positive velocity structure.
    The image is oriented with north up and east to the left. 
    With black we note the logarithmically scaled \textit{XMM-Newton} count contours derived from the combined data set of Fig.~\ref{fig:XMM_img}. 
    Black arrows indicate three elongated soft X-ray features.
    One of them coincides with the direction of the northern outflow cone.
    With a white box, the extent of Fig.~\ref{fig:Chandra_img} is indicated, highlighting the difference in scale between the diffuse X-ray emission in the halo and the extent of the galaxy.  }
    \label{fig:Havel_img}
\end{figure}

Optical observations reveal that ESO 338-4 drives galactic outflows, traced by \element{H}$\alpha$ velocity measurements.
In particular, a prominent outflow cone has been linked to CL23, a bright stellar cluster located at the base of the structure and surrounded by an open superbubble \citep{Bik2018}. 
These outflows suggest significant feedback activity capable of shaping the circumgalactic medium. 

Our new deep \textit{XMM-Newton} observations allow us to map the distribution of hot gas and compare it to the morphology seen in optical.
In Fig.~\ref{fig:Havel_img}, we present the \element{H}$\alpha$ velocity field from \citet{Bik2018} overlaid with X-ray count rate contours derived from the combined exposures shown in Fig.~\ref{fig:XMM_img}.
This comparison highlights the spatial distribution of the diffuse X-ray emission relative to the galaxy’s optical morphology.
The X-ray contours reveal an elongated structure extending northward, spatially coincident with the direction of the optical outflow. 
This soft X-ray feature may be linked to the galaxy-wide feedback processes, and potentially to the activity of CL23 and its associated superbubble.
However, the superbubble itself remains undetected in X-rays by \textit{Chandra}, likely because it is too faint to be distinguished from the surrounding bright ULXs.

\begin{figure*}[h!]
\sidecaption
        \centering
        \includegraphics[width=12cm]{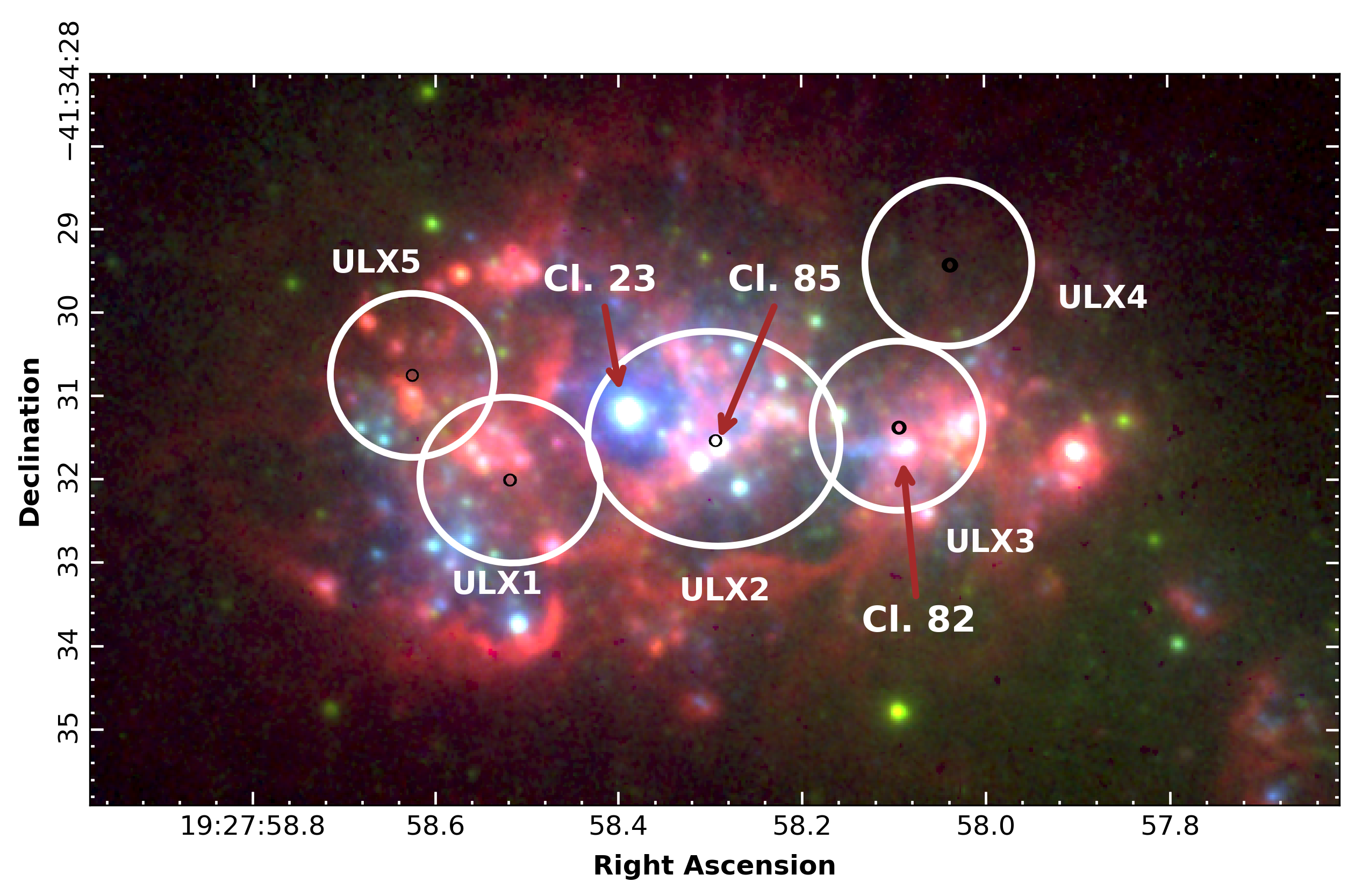}
        \caption{Three-color composite HST image of ESO 338-4. Red is centered on the H$\alpha$ emission line with the F656N filter, blue is in UV with the F140LP filter, and green is the F550M filter. The ULX point source positions (Table~\ref{tab:ULX_prop}) along their $1\sigma$ uncertainties are
overlaid on this image in black.
We highlight the boundaries of the \textit{Chandra} count distribution of each source at a 3$\sigma$ level through white ellipses (Fig.~\ref{fig:Chandra_img}). An association between ULX2 and CL85, and ULX3 and CL82 is established as their positional uncertainty is within the optical extent of the clusters.}
        \label{fig:HST_img}
\end{figure*}

\subsubsection{Hot gas mass}

Based on the spectral fits presented in Sect.~\ref{sec:spec_anal}, we estimate the mass of the hot gas in ESO 338-4. 
This is done using the emission measures ($EM$) of the halo’s diffuse emission components, defined as $EM = \int n_e n_H dV$, where $n_e$ and $n_H$ are the electron and hydrogen densities, respectively, and $dV$ is the volume element. 
In the \texttt{APEC} model, the thermal plasma normalization relates to the emission measure via $norm=\frac{10^{-14}}{4\pi D^2}\times EM$, assuming negligible redshift and a distance $D$ to the galaxy. 
Plasma models in \texttt{XSPEC} assume a fully ionized gas with a fixed ratio of $n_e/n_H = 1.2$. This ratio is accurate to within $3.5\%$ for solar-abundance plasmas at temperatures between $0.10$ and $40$\,keV \citep{xspecratio2024}.
Given the low metallicity of ESO 338-4 and the fact that deviations from this ratio are smaller than the uncertainties on the normalization of the \texttt{APEC} component, the assumed ratio of 1.2 remains an appropriate approximation.

Potential metal enrichment in the halo could imply different best-fit parameters of the plasma models.
To test for deviations from the previously assumed metallicity of $Z = 0.12\,Z_\odot$, we repeated the spectral fitting using two \texttt{VAPEC} components in place of the two \texttt{APEC} models. 
The best-fit scenario returns an enrichment of oxygen and neon of $Z_{\rm O}=(0.2\pm 0.1)\,Z_\odot$ and $Z_{\rm {Ne}}=(1.0\pm 0.4)\,Z_\odot$ with the other element abundances fixed at $0.12\,Z_\odot$.
The resulting plasma temperatures and normalizations are consistent, within uncertainties, with those obtained using the original \texttt{APEC} models.
Therefore, for consistency and simplicity, we proceed with the parameters listed in Table~\ref{tab:Spec_pam} for all further calculations.

Estimating the halo’s hot gas mass requires assumptions about both the density profile and the geometry of the emitting volume.
For the former, we adopt a homogeneous distribution of matter, while for the latter, we assume a sphere.
The optical extent of the galaxy (Fig.~\ref{fig:XMM_img}) is excluded from this volume and is modeled separately as a cylinder with a height equal to the sphere’s radius.
We derive $M^{\rm halo,soft}_{\ion{H}{ii}}=(4.4\pm0.3)\times\,10^7\,\text{M}_\odot$ for the $0.26$\,keV ($T_1=3.0$\,MK) and $M^{\rm halo,hard}_{\ion{H}{ii}}=(6.8\pm0.1)\times 10^7\,\text{M}_\odot$ for the  $3.5$\,keV  ($T_2=4$\,MK) component. 
Performing the same calculation for the diffuse component of the cylindrical central region yields $M^{\rm central}_{\ion{H}{ii}}=(7.03\pm0.09)\times 10^7\,\text{M}_\odot$.
Thus, based on \textit{XMM-Newton} observations, we estimate the total hot gas mass in ESO 338-4 as $M^{\rm diff}_{\ion{H}{ii}}=(18.2\pm0.4)\times10^7\,\text{M}_\odot$, with the halo alone contributing $M^{\rm halo}_{\ion{H}{ii}}=(11.2\pm0.4)\times10^7\,\text{M}_\odot$.

Our hot gas mass estimate is a few times higher than that derived from optical data. \citet{Bik2018} report the H$\alpha$ emitting gas mass as $M_{\text{\ion{H}{ii}}} = 3.0 \times 10^7\,\text{M}_\odot$ using a modeled density profile for the galaxy. In contrast, an earlier warm gas mass estimate assuming a constant density yielded $M_{\text{\ion{H}{ii}}} = 16 \times 10^7\,\text{M}_\odot$ \citep{Oldhalomass1999}.
These comparisons suggest that the hot and warm ionized gas masses in ESO 338-4 are broadly similar.
All ionized mass estimates are however, significantly lower than the neutral hydrogen mass of $M_{\text{\ion{H}{i}}}=(1.4 \pm 0.2) \times 10^9\,\text{M}_\odot$ suggested by \cite{Neutralmass2004} utilizing \ion{H}{i} observations.

\subsubsection{X-ray Luminosity}

The soft X-ray emission originates from the interaction of the large amounts of mechanical energy released by the starburst episode with the interstellar medium around the newly formed stars, as evidenced by the hot bubbles identified around the central clusters of ESO 338-4 in the optical images. 
The total release of mechanical energy from stellar winds and supernova explosions has been computed in several studies following the prescriptions of \citet{Leitherer1992}, including \citet{Cervi2002}, \citet{Stevens2003}, \citet{Oskinova2005}, and \citet{otifloranes2010}. 
As concluded by these authors, the soft X-ray luminosity of star-forming galaxies can be reproduced by their synthesis models assuming the average fraction of mechanical energy being released in the form of soft X-rays by the heated gas is in the range $X_{\rm eff}=1\text{--}10$\%. 
Using the calibrations by \cite{otifloranes2010}, a  $\log(L_X\,[\text{erg}~\text{s}^{-1}]) = 40.26$ originated in the diffuse medium with an SFR of $3.3~\text{M}_\odot\,\text{yr}^{-1}$ over more than $30$\,Myr, would require  a realistic efficiency $X_{\rm eff}\sim 0.02$ (assuming an SFR computed for a Salpeter initial mass function within a mass range $0.1\text{--}100\,M_\odot$). This would confirm  that the observed diffuse soft X-ray emission is consistent with being originated from the mechanical energy released by the massive stars after  interaction with the interstellar medium.

We compare the X-ray halo luminosity of ESO 338-4 with that of other dwarf starburst galaxies.
In \cite{Otherdwarfhalos2005}, seven galaxies have been analyzed. 
Five of them (NGC 3077, NGC 4449, NGC 5253, NGC 4214, He 2-10) have been found to feature X-ray bright galactic halos, with two non-detections (VII Zw 403, I Zw 18).
Their analysis discusses a positive correlation between the halo luminosity and host galaxy metallicity.

Among the X-ray bright halos, NGC 1569 exhibits the lowest luminosity, $L^{0.3\text{--}8\,\text{keV}}_X=4.2 \times 10^{38}$~erg s$^{-1}$, and a galaxy metallicity of $12+\log(\element{O}/\element{H})=8.22$. 
The brightest halo is found in He~2--10\footnote{In a later study, He~2--10 was found to host an accreting supermassive black hole \citep{He211_BH_2011}.}, with $L^{0.3\text{--}8\,\text{keV}}_X=2.0 \times 10^{40}$~erg s$^{-1}$ and a near-solar metallicity. 
A comparable halo luminosity is measured in ESO~338-4, $L^{0.3\text{--}8\,\text{keV}}_X=1.68 \times 10^{40}$~erg s$^{-1}$. 
The galaxy, however, is a dwarf starburst with a metallicity of $12+\log(\element{O}/\element{H})=7.9$, lower than any of the five galaxies with hot gas detections in their sample. 
Interestingly, the SFR of ESO 338-4 is higher than that of any system in their sample. 
Therefore, we suggest that the halo $L_X$ correlates primarily with the SFR rather than $Z$, a relation that has likewise been reported for spiral galaxies \citep{Strickland2004b,LiWang2013}

\subsection{X-ray sources as sources of \ion{He}{ii} ionizing radiation}\label{sec:cloudy}

To quantify the importance of X-rays to the ionizing photon budget of ESO 338-4, we first need to search for optical counterparts to the ULXs.
The spatial separation between ULXs and clusters is particularly important when discussing the distinct components of the narrow and broad \ion{He}{ii}\,$\lambda 4686$\,\AA~emission: while the broad component originates from WR stars, the narrow emission arises from nebulae ionized either by massive stars or X-ray sources.
When WR stars and ULXs are spatially coincident, the combined contribution to the nebular ionizing flux complicates the interpretation of the narrow component. Moreover, the inherent difficulty of isolating the narrow emission adds an additional challenge to studies of nebular photoionization.

In ESO 338-4, four clusters, CL23, 18, 53, and 50/51\footnote{Resolved initially by \cite{Ostlin98} as two separate clusters but identified as one with new fitting \citep{Bik2018}} exhibit WR features. However, none of the detected ULXs coincide with these WR clusters.
Some ULXs are, nonetheless, spatially associated with other stellar clusters. 
Using the cluster IDs of \cite{Ostlin98}, we confirm the association between ULX2 and CL85, and additionally identify ULX3 as coincident with CL82.
This spatial distribution is illustrated in Fig.~\ref{fig:HST_img}, which shows the ULX positions overlaid on a color composite of archival HST data.

Recently, \cite{KourSv2025} derived an empirical relation between the observed \ion{He}{ii}\,$\lambda 4686$\,\AA~emission and the X-ray luminosity for a sample of 165 galaxies from the Chandra Source Catalogue. Our measurements are consistent with their best-fit relation for star-forming galaxies to within $1\sigma$, suggesting that the ULX properties of ESO 338-4 do not require fine-tuning to account for the observed \ion{He}{ii}\,$\lambda 4686$\,\AA~emission. The galaxy behaves consistently with the general population of \ion{He}{ii} and X-ray emitters.

To estimate the contribution of X-ray sources to the ionization budget of ESO 338-4, we use photoionization modeling of the time-averaged ULX1 emission and compare it to the observed \ion{He}{ii}\,$\lambda 4686$\,\AA~emission.
For this prediction, calculations were performed with version 23.01 of \textit{Cloudy} \citep{Cloudy2023,Cloudy2}.
We assume a spherical geometry and a hydrogen density of $\log({n_{\element{H}}[\text{cm}^{-3}]})=2$.
\textit{Cloudy} requires both an input SED and a luminosity defined over a specified energy interval, with the latter used to set the overall normalization.
For the SED, we adopt the best-fit model of ULX1 (Table~\ref{tab:Spec_pam}) corrected for Galactic and intrinsic absorption.
Choosing the range of $0.2$--$10$ keV for the normalization, i.e., the range of our \textit{XMM-Newton} data, the absorption-corrected (intrinsic and galactic) luminosity of ULX1 is $\log(L_X[\text{erg}~\text{s}^{-1}])=41.4$.
This results in a predicted nebular \ion{He}{ii}\,$\lambda 4686$\,\AA~luminosity of $L_{\text{\ion{He}{ii}}~4686,\text{nebular}}=1.9\times 10^{39}\text{erg}~\text{s}^{-1}$ comparable to the total (narrow+broad) observed value of $L^{\text{obs}}_{\ion{He}{ii}~4686,\text{total}}\approx2\times 10^{39}\text{erg}~\text{s}^{-1}$ \citep{LidaPap2019}.

Our calculation does not include the stellar contribution from WR stars, as our aim here is specifically to isolate the potential impact of the ULX on the excitation of \ion{He}{ii}.
A detailed photoionization model incorporating the combined effects of the WR clusters and the ULX is beyond the scope of the present study and will be addressed in future work.
Our estimate implicitly assumes that the ULX radiation escapes isotropically and without significant attenuation by circumsource material, and therefore represents an upper limit to the resulting nebular \ion{He}{ii} luminosity.

While the very high X-ray luminosity by ULX1 would therefore be enough to explain the observed nebular \ion{He}{ii} luminosity, the ionizing capability of the other ULXs wouldn't be clearly enough, with the associated $L_{\ion{He}{ii}~4686}$ values a factor above approximately 5 lower than the observed one. We therefore suggest looking for very bright ULXs in star-forming galaxies with an excess of $L_{\ion{He}{ii}~4686}$ emission. 

On the other hand, the X-ray luminosity of the other ULXs in ESO338-4 is similar to the integrated X-ray luminosity of the diffuse gas, which originates from gas heated by the release of mechanical energy from massive stars. While the extrapolation to the far UV of the ULX continua is not straightforward, the diffuse intrinsic X-ray emission usually peaks below $0.5\text{--}2$\,keV, providing a significant amount of photons able to ionize \ion{He}{ii}, as proposed by \cite{Cervi2002} using a Cloudy modeling similar to ours. Therefore, we conclude that in the absence of very bright ULXs, the \ion{He}{ii} emission could originate from soft X-ray photons produced by lower luminosity ULXs and/or hot, diffuse gas \citep{OskinovaScharer2022}.     

\section{Conclusions}\label{sec:Conclusions}

We determined and presented the X-ray properties of the dwarf starburst galaxy ESO 338-4 based on new deep observations with \textit{Chandra} and \textit{XMM-Newton}. Within the optical extent of the galaxy, we detect five point sources, all classified as ULXs with $L_X>10^{39}\,\text{erg s}^{-1}$. The brightest, ULX1, is strongly variable on a timescale of days, with luminosity changes exceeding one order of magnitude. No coherent pulsations are detected in the central region in the \textit{XMM-Newton} data. Two ULXs are found in the vicinity of massive stellar clusters, while the others — including ULX1 — are not directly associated with any observed cluster at the 1$\sigma$ level.

The total point source luminosity in the galaxy is $\log(L_X[\text{erg}~\text{s}^{-1}])=40.8$ when corrected for Galactic absorption, or $\log(L_X[\text{erg}~\text{s}^{-1}])=41.3$ when including corrections for the intrinsic absorption. Assuming a SFR of $0.6$\,M$_\odot$\,yr$^{-1}$, these values correspond to $\log(L_X/\text{SFR}~[\text{erg}~\text{s}^{-1}(\text{M}_\odot\,\text{yr}^{-1})^{-1}])=41.1$ and $41.5$, respectively, highlighting the high X-ray output per unit star formation rate in ESO 338-4.

ESO 338-4 also exhibits bright diffuse X-ray emission extending far beyond the optical extent of the galaxy. This emission is well described by a two-temperature plasma model with $T_1=3.0$\,MK ($0.26$\,keV) and $T_2=4$\,MK ($3.5$\,keV).
The X-ray emission from the diffuse gas is consistent with predictions from synthesis models, which indicate that it is produced by the release of mechanical energy from massive stars, with an efficiency of $X_{\rm eff}\approx 0.05$ in converting mechanical to thermal energy, which is then re-emitted in this range.  
The total X-ray luminosity of the hot halo is $L^{0.3-8\text{keV}}_X=1.7 \times 10^{40}$~erg s$^{-1}$, and the corresponding hot gas mass is $M^{\rm halo}_{\ion{H}{ii}}=(11.2\pm0.4)\times10^7\,\text{M}_\odot$. Its morphology shows a northern elongation aligned with the direction of the galactic outflow traced by H$\alpha$ kinematics.

Finally, we assess the contribution of ULX1 to the galaxy's ionizing photon budget. Using spectral modeling of the unabsorbed time-averaged ULX1 photon flux, we predict a high nebular \ion{He}{ii}\,$\lambda 4686$\,\AA~luminosity of  $L^{\text{obs}}_{\text{HeII}4686,\text{total}}\approx2\times 10^{39}\text{erg}~\text{s}^{-1}$. 
This value is comparable to the combined broad and narrow (corresponding to stellar and nebular components, respectively) \ion{He}{ii}\,$\lambda 4686$\,\AA~luminosity observed in ESO 338-4.
This shows that X-rays, and ULXs in particular, can be significant contributors to the hard ionizing radiation field in low-metallicity starburst systems. The total modeled X-ray luminosity of the galaxy in the $0.054\text{--}12$\,keV band is $\log(L^{\text{model}}_{0.054-12\text{\,keV}}$[erg s$^{-1}])=41.6$, with ULX1 alone contributing $\log(L^{\text{model}}_{0.054-12\text{\,keV}}$[erg s$^{-1}])=41.5$.

\begin{acknowledgements}
We thank the anonymous referee for the helpful comments,
which have improved the quality of this paper.
The scientific results reported in this article are based to a significant degree on observations made by the Chandra X-ray Observatory.
This work has made use of data from the European Space Agency (ESA) mission
    {\it Gaia} (\url{https://www.cosmos.esa.int/gaia}), processed by the {\it Gaia}
    Data Processing and Analysis Consortium (DPAC,
    \url{https://www.cosmos.esa.int/web/gaia/dpac/consortium}). Funding for the DPAC
    has been provided by national institutions, in particular the institutions
    participating in the {\it Gaia} Multilateral Agreement.
    This research has made use of the SIMBAD database, operated at CDS, Strasbourg, France, and the NASA's Astrophysics Data System Bibliographic Services.
    The research presented in this paper is funded by the {\it Deutsche Forschungsgemeinschaft} (DFG, German Research Foundation) -{\it Projektnummer} 529885128.
    JMMH is funded by Spanish MICIU/AEI/10.13039/501100011033 and ERDF/EU grant PID2023-147338NB-C21.
     M.J.H. is supported by the Swedish Research Council (Vetenskapsr{\aa}det) and is Fellow of the Knut \& Alice Wallenberg Foundation.
     Partial support for JSG's participation in this work was provided by the National Aeronautics and Space Administration through Chandra Award Number GO2-203062X issued by the Chandra X-ray Center, which is operated by the Smithsonian Astrophysical Observatory for and on behalf of the National Aeronautics Space Administration under contract NAS8-03060.
     The Authors thank Angela Adamo for their feedback at the initial stages of this work.
\end{acknowledgements}

\bibliographystyle{aa} 
\bibliography{refs}

\begin{appendix}

\section{Log of observations}\label{app:log}
\begin{table}[h!]
\caption{Log of Observations of ESO 338-4 used in this paper.}                
\centering                       
\begin{tabular}{c c c}     
\hline\hline               
Observation ID & Exposure time (ks) & Start date  \\         
\hline
\noalign{\smallskip}
\multicolumn{3}{c}{\textit{Chandra} (Proposal Nr. 23620207)}\\
\hline
\noalign{\smallskip}
25236 	& 14.90  & 28 July 2023\\
25768 	& 13.90 & 18 July 2023  \\
25769& 11.88 & 29 June 2023 \\
25770& 14.89 	  & 22 July 2023 \\
25771 & 11.93 & 7 July 2023 \\
25772 	& 10.45 & 4 April 2023 \\
25773 	& 9.96 & 5 April 2023 \\
25774& 17.74 	 & 6 April 2023 \\
25775 	& 31.67 &15 April 2023 \\
25776& 29.70 & 28 September 2023 \\
27743& 16.86 & 17 March 2023 \\
27791& 49.43 & 9 April 2023 \\
27925& 15.88 & 30 June 2023 \\
27941 	& 16.37 	 &8 July 2023\\
27959 	& 14.89 &23 July 2023 \\
27964& 14.90 & 28 July 2023\\
\hline
\noalign{\smallskip}
\multicolumn{3}{c}{\textit{XMM-Newton}}\\
\hline
0892410101 &  90.3 & 17 October 2021\\
0780790201 & 24.001 & 10 April 2016\\
\noalign{\smallskip}
\hline
\end{tabular}
\end{table}

\section{The extended halo}\label{app:extended_halo}
\begin{table}[h!]
\caption{
Best-fit spectral parameters for the extended halo region of ESO~338-4 from the \textit{XMM-Newton} data. 
}                    
\centering                       
\begin{tabular}{l  c}     
\hline\hline               
Model Parameter & Best-fit value  \\        
\hline
\noalign{\smallskip}
\multicolumn{2}{c}{\texttt{tbabs$_{\texttt{Gal}}$} $\times$ (\texttt{APEC}$_1$ + \texttt{APEC}$_2$)}\\
\noalign{\smallskip}
\hline
\noalign{\smallskip}
kT$_1$ (keV) & $0.4\pm0.1$\\
Norm$_1$ (cm$^{-3}$)& $(8\pm3)\times10^{-6}$\\
kT$_2$ (keV) &$4\pm1$\\
Norm$_2$ (cm$^{-3}$)& $(2.0\pm0.2)\times10^{-5}$\\
C-statistic &  178\\
d.o.f & 139\\
$p_{\rm null}$ & 0.021\\
\hline
\noalign{\smallskip}
\multicolumn{2}{c}{\texttt{tbabs$_{\texttt{Gal}}$} $\times$ (\texttt{APEC}$ + $ \texttt{const} $\times$ (Point-source contribution))}\\
\noalign{\smallskip}
\hline   
\noalign{\smallskip}
kT (keV)& $0.32\pm0.04$ \\
Norm (cm$^{-3}$)& $(2.0\pm0.3)\times10^{-6}$\\
Factor & $(5.8\pm0.4)\times10^{-2}$ \\
C-statistic & 177\\
d.o.f & 140\\
$p_{\rm null}$ & 0.026\\
 \hline
\end{tabular}
\tablefoot{Two alternative models are shown: (i) a two-component thermal plasma (\texttt{APEC}$_1$+\texttt{APEC}$_2$) and (ii) a single plasma component plus a point-source contamination term. 
Both models yield statistically comparable fits, consistent with the analysis presented in Sect.~\ref{sec:spec_anal}. 
The Galactic foreground absorption was fixed to $N_{\mathrm{H}} = 5\times10^{20}\,\mathrm{cm}^{-2}$ using \texttt{tbabs}$_{\mathrm{Gal}}$.}
\end{table}

\newpage

\section{Chandra Variability of ULX2--5}\label{app:var}

\begin{figure}[h!]
\centering
\includegraphics[width=\hsize]{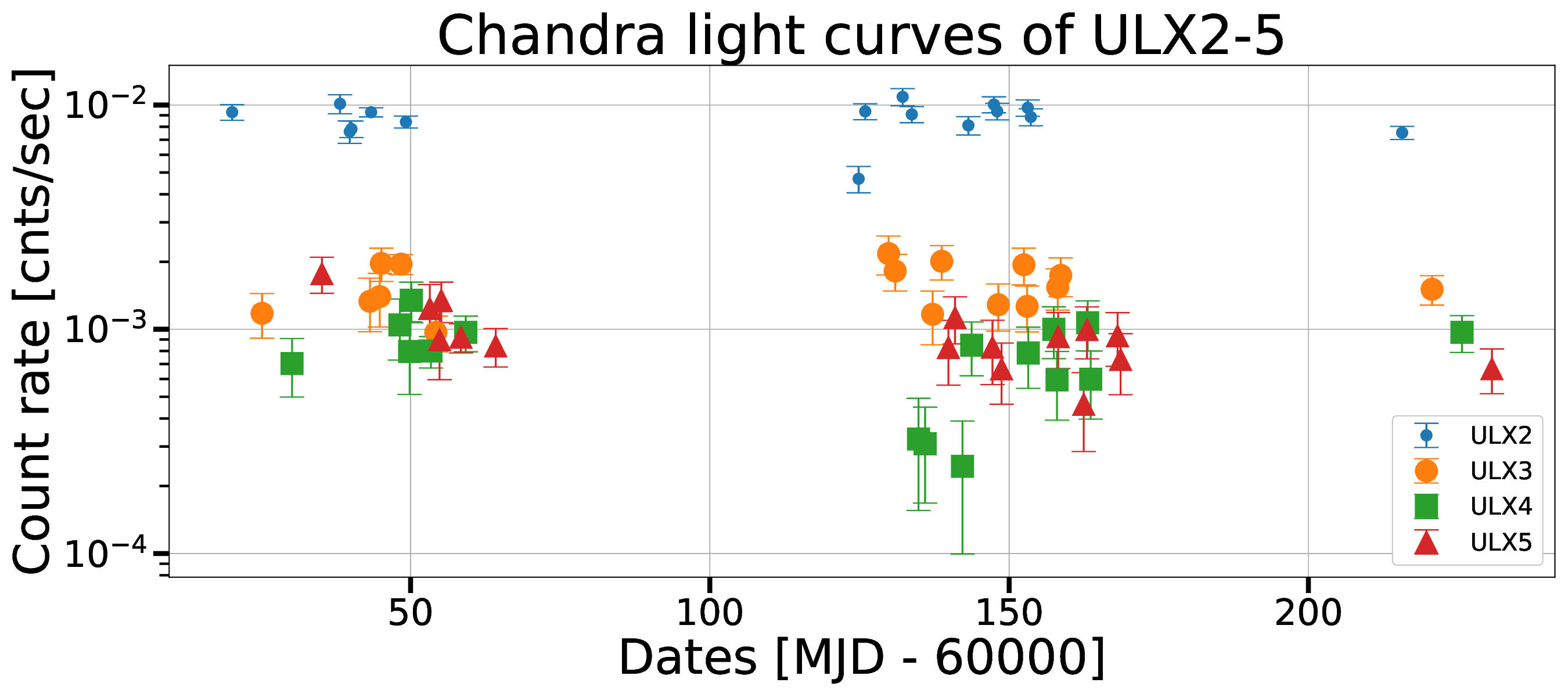}
  \caption{Upper panel: Light curves of ULX2--5 from the 16 \textit{Chandra} observations obtained in 2023. The data points are constructed from the total count rates in the $0.5$--$7$\,keV range of each respective observation. To improve visual clarity, we applied successive $+$5-day offsets to the observation dates of ULX3--5, with ULX2 left unshifted. }
\end{figure}

\section{Chandra spectra}\label{app:CH_spec}

\begin{figure*}[h!]
    \centering
 
    \begin{subfigure}[t]{0.48\textwidth}
        \centering
        \includegraphics[trim=0 0 10 142, clip,angle=-90,width=\hsize]{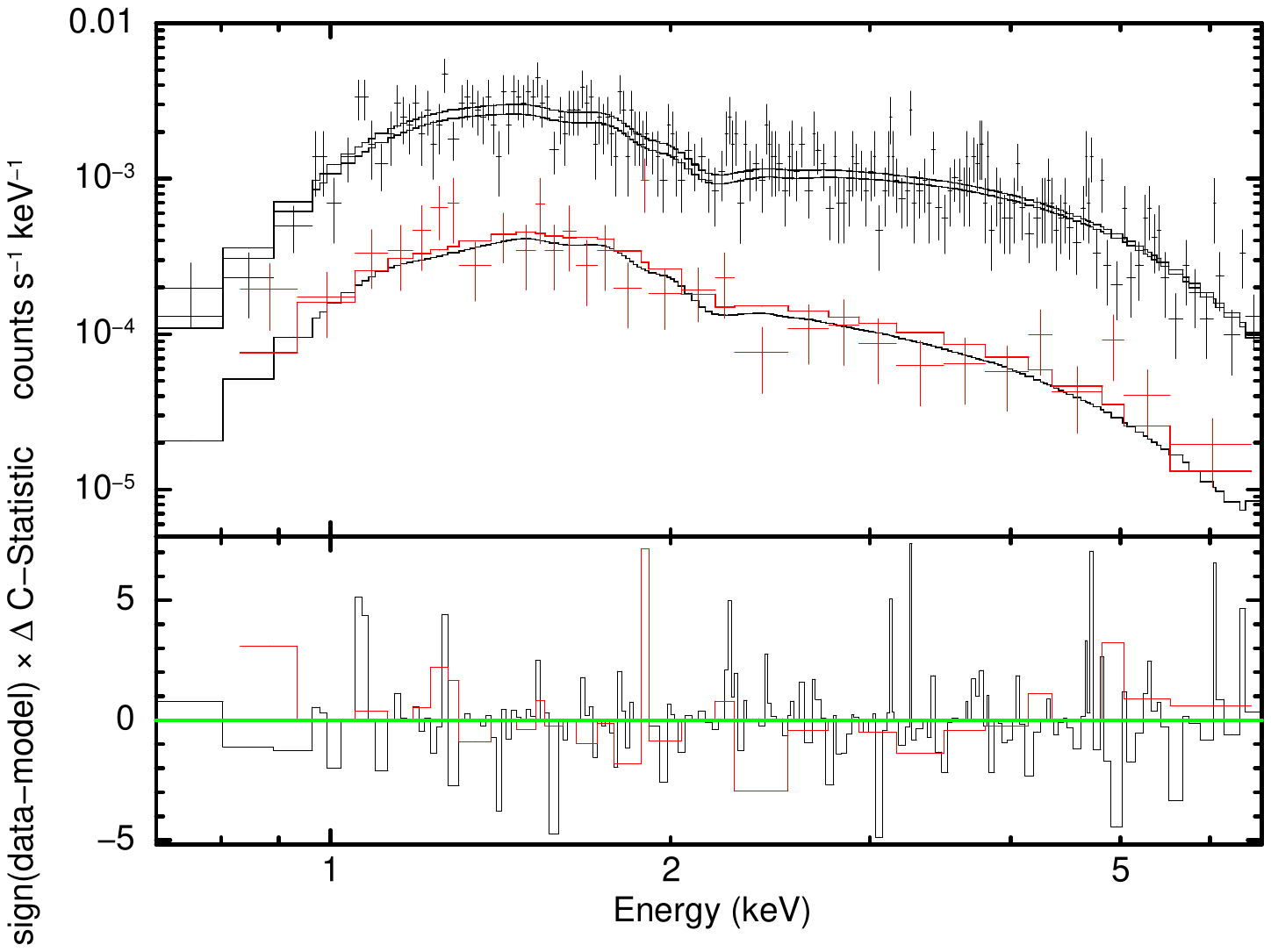}
        \caption{ULX1}
    \end{subfigure}
    \hfill
    \begin{subfigure}[t]{0.48\textwidth}
        \centering
        \includegraphics[trim=0 0 10 142, clip,angle=-90,width=\hsize]{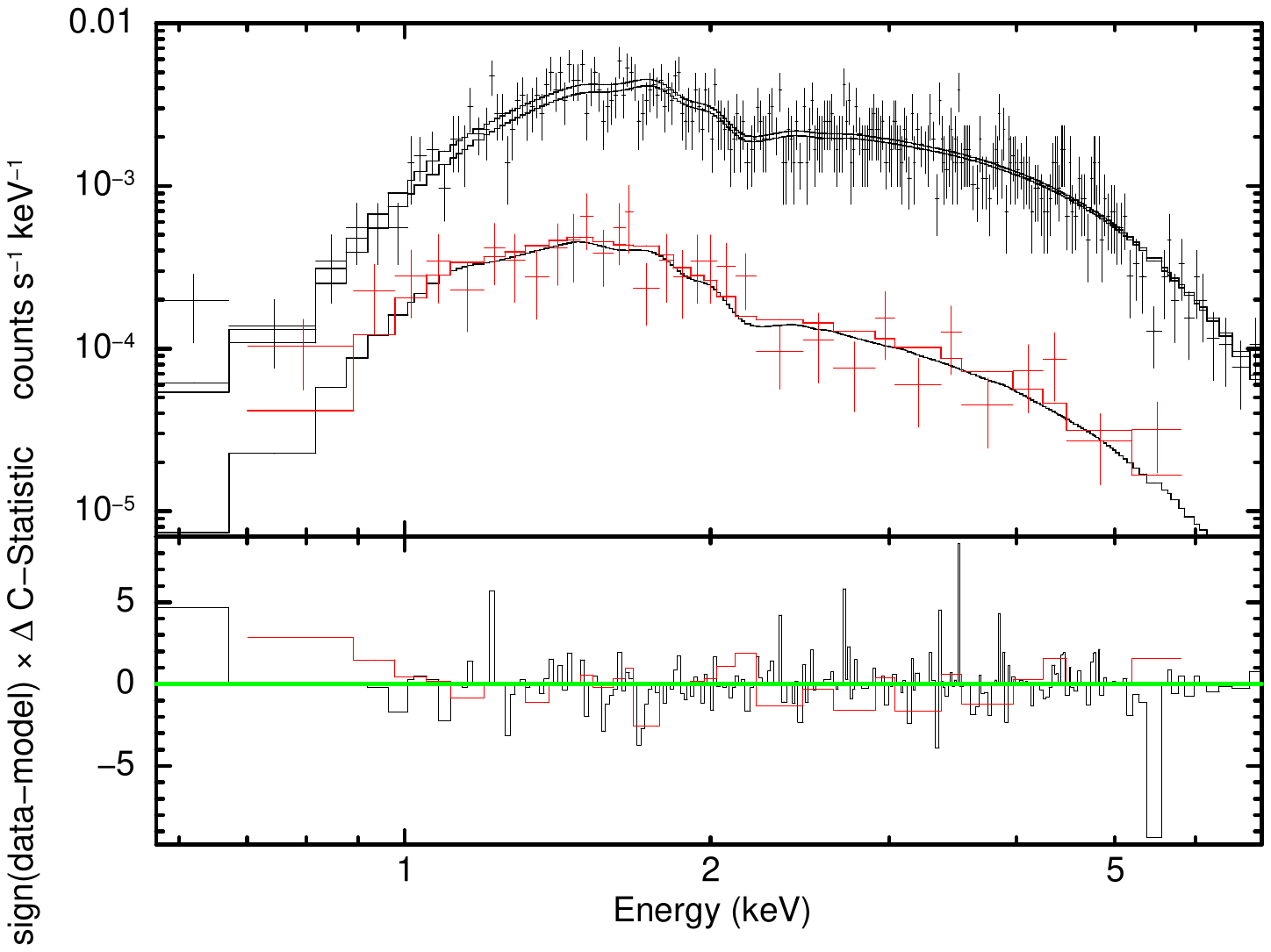}
        \caption{ULX2}
    \end{subfigure}
    \hfill
    \begin{subfigure}[t]{0.48\textwidth}
        \centering
        \includegraphics[trim=0 0 10 142, clip,angle=-90,width=\hsize]{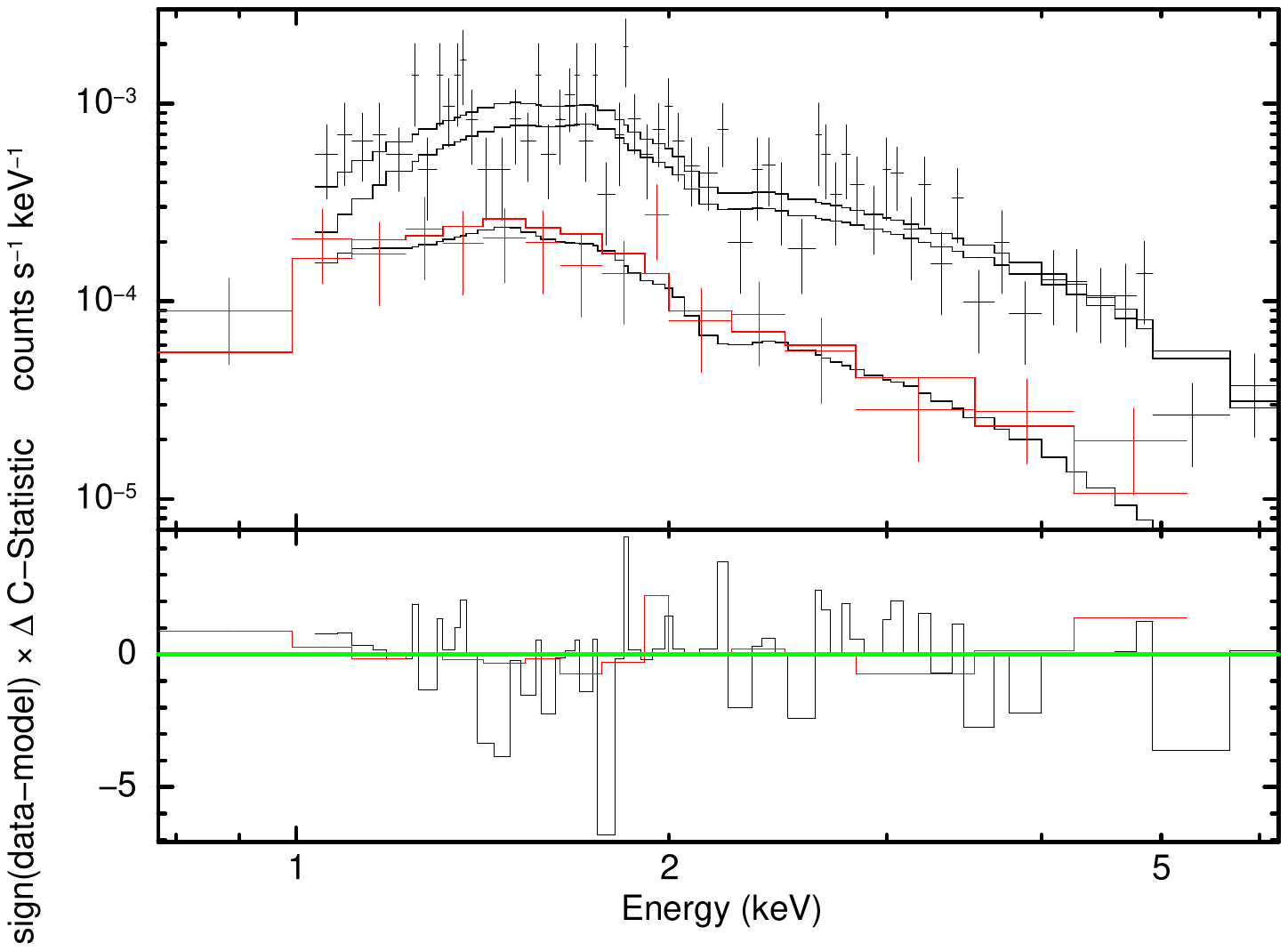}
        \caption{ULX3}
    \end{subfigure}
    \hfill
    \begin{subfigure}[t]{0.48\textwidth}
        \centering
        \includegraphics[trim=0 0 10 142, clip,angle=-90,width=\hsize]{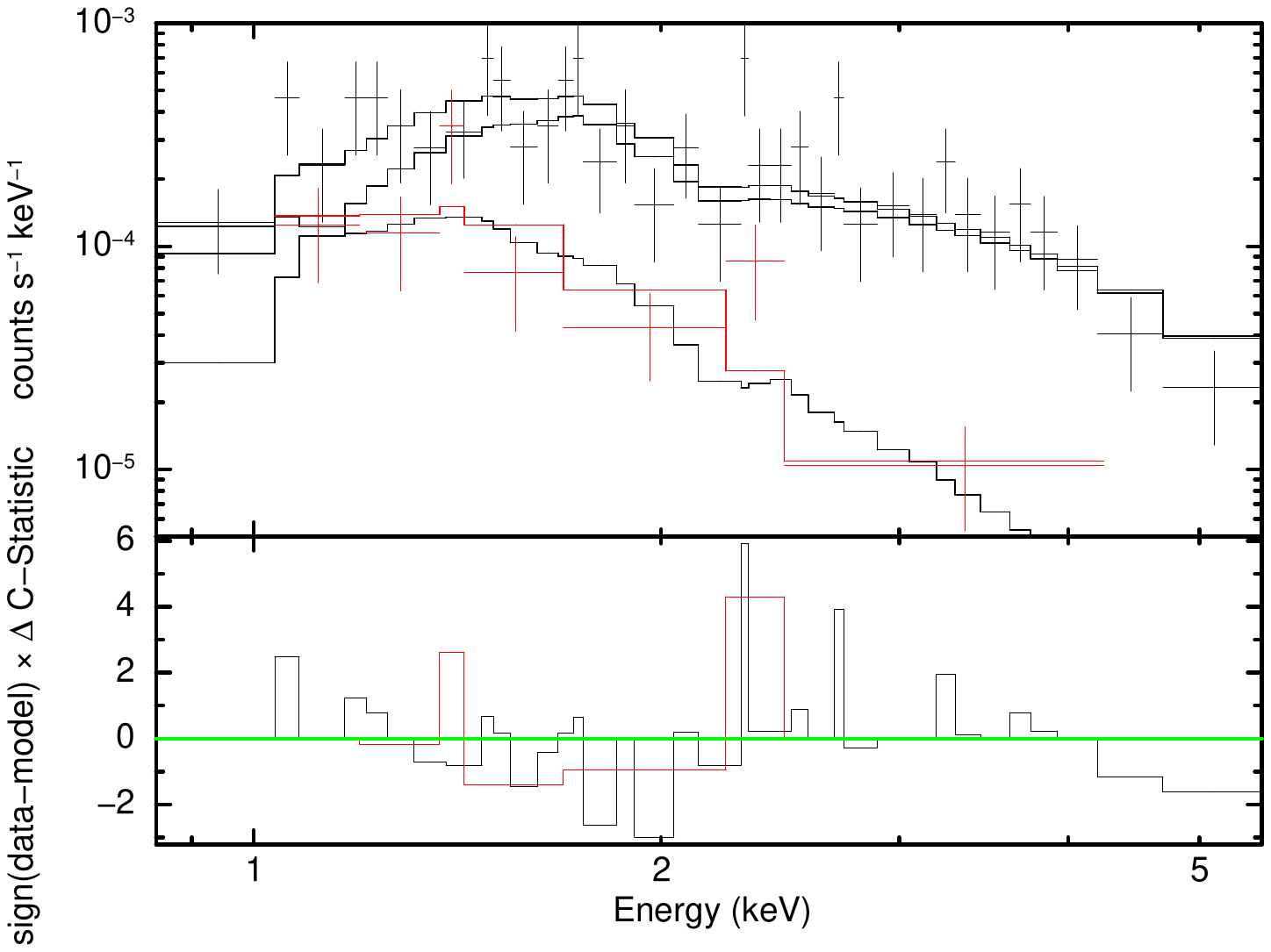}
        \caption{ULX4}
    \end{subfigure}
    \hfill
    \begin{subfigure}[t]{0.48\textwidth}
        \centering
        \includegraphics[trim=0 0 10 142, clip,angle=-90,width=\hsize]{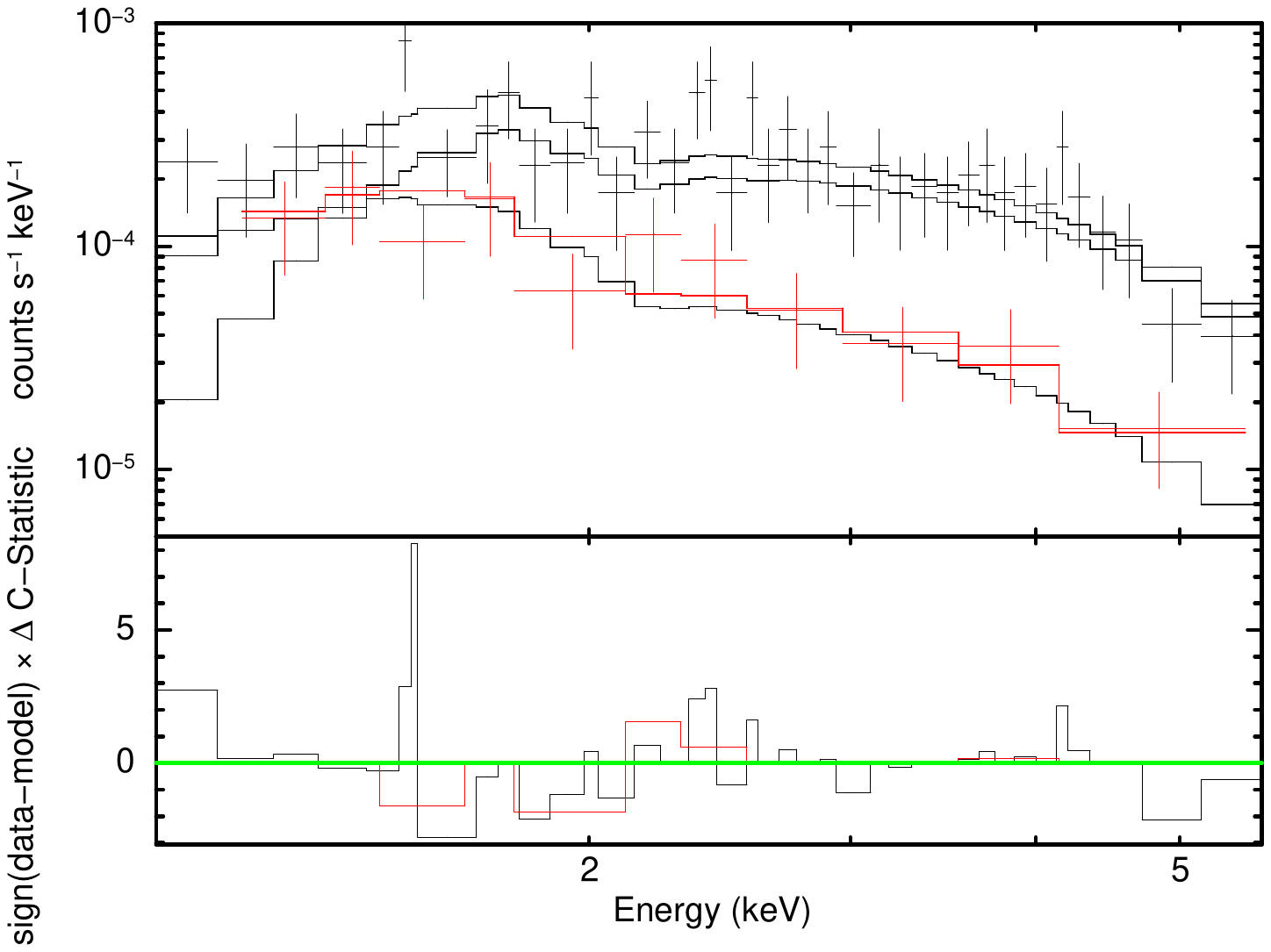}
        \caption{ULX5}
    \end{subfigure}

    \caption{\textit{Chandra} spectra. Simultaneous \texttt{xspec} fit of the 5 point sources (black) and their local diffuse background (red). The C-statistic residuals of each fit are found in the bottom panel of each figure. A detailed discussion on the spectral fits and \texttt{xspec} models is found in Sect.~\ref{sec:spec_anal}}
    \label{fig:appendix_3x2_layout}
\end{figure*}

\end{appendix}

\end{document}